d
\documentclass[aoas,MSNbibl,nameyear,dvips]{arximspdf}
\usepackage{dcolumn}
\usepackage{subenv}
\usepackage[ruled,vlined]{algorithm2e}
\usepackage{mathbh}
\usepackage{tikz}
\usepackage{graphicx}

\doi{10.1214/11-AOAS520}
\volume{6}
\issue{2}
\pubyear{2012}
\firstpage{795}
\lastpage{830}

\makeatletter
\newcolumntype{d}[1]{D{.}{.}{#1}}

\newtheorem{theorem}{Theorem}
\newtheorem{proposition}{Proposition}
\newtheorem{lemma}{Lemma}

\newcommand{\argmax}{\operatorname{\arg\max}\limits}
\newcommand{\argmin}{\operatorname{\arg\min}\limits}
\newcommand{\Rset}{\mathbb{R}}
\newcommand{\norm}[1]{\|#1\|}
\newcommand{\group}[1][k]{\mathcal{G}_{#1}}
\newcommand{\1}{\mathbh{1}}
\newcommand{\bbeta}{\boldsymbol\beta}
\newcommand{\bvarphi}{\boldsymbol\varphi}
\newcommand{\btheta}{\boldsymbol\theta}
\newcommand{\bPsi}{\boldsymbol\Psi}
\newcommand{\sign}{\operatorname{sign}}
\newcommand{\supp}{\mathcal{S}}
\newcommand{\prob}{\mathbb{P}}
\newcommand{\inprob}{\stackrel{P}{\longrightarrow}}
\newcommand{\lambdaclass}{\lambda^{\mathrm{err}}}
\newcommand{\hatbeta}{\hat{\beta}}
\newcommand{\hatbbeta}{\hat{\bbeta}}
\newcommand{\hatbetacoop}{\hat{\beta}{}^{\mathrm{coop}}}
\newcommand{\hatbetaridge}{\hat{\beta}{}^{\mathrm{ridge}}}
\newcommand{\hatbbetalasso}{\hat{\bbeta}{}^{\mathrm{lasso}}}
\newcommand{\hatbbetacoop}{\hat{\bbeta}{}^{\mathrm{coop}}}
\newcommand{\hatbbetagroup}{\hat{\bbeta}{}^{\mathrm{group}}}
\newcommand{\hatbbetaridge}{\hat{\bbeta}{}^{\mathrm{ridge}}}
\newcommand{\hatbbetaols}{\hat{\bbeta}{}^{\mathrm{ols}}}
\newcommand{\tildebbeta}{\tilde{\bbeta}{}}
\newcommand{\bbetaols}{\bbeta{}^{\mathrm{ols}}}

\newcommand{\eqref}[1]{(\ref{#1})}
\newcommand{\cal}{\mathcal}
\makeatother

\begin{document}
\begin{frontmatter}

\title{Sparsity with sign-coherent groups of variables via the
cooperative-Lasso}
\runtitle{Sparsity with sign-coherent groups of variables}

\begin{aug}
\author[A]{\fnms{Julien} \snm{Chiquet}\corref{}\ead[label=e1]{julien.chiquet@genopole.cnrs.fr}},
\author[A]{\fnms{Yves} \snm{Grandvalet}\thanksref{au2}\ead[label=e2]{yves.grandvalet@utc.fr}}
\and
\author[A]{\fnms{Camille} \snm{Charbonnier}\ead[label=e3]{camille.charbonnier@genopole.cnrs.fr}\ead[label=u,url]{http://stat.genopole.cnrs.fr}}
\runauthor{J. Chiquet, Y. Grandvalet and C. Charbonnier}
\affiliation{CNRS UMR 8071 \& Universit\'e d'\'Evry and
Universit\'e de Technologie de~Compi\`egne---CNRS UMR 6599 Heudiasyc}
\address[A]{Laboratoire Statistique et G\'enome\\
23, boulevard de France\\
91037 \'Evry\\
 France\\
\printead{e1}\\
\hphantom{\textsc{E-mail:} }\printead*{e2}\\
\hphantom{\textsc{E-mail:} }\printead*{e3}\\
\printead{u}} 
\end{aug}
\thankstext{au2}{Supported in part by the PASCAL2 Network of Excellence,
the European ICT FP7   Grant   247022---MASH and the French National
Research Agency (ANR)   Grant ClasSel ANR-08-EMER-002.}

\received{\smonth{3} \syear{2011}}
\revised{\smonth{9} \syear{2011}}

%
\begin{abstract}
We consider the problems of estimation and selection of parameters endowed
with a known group structure, when the groups are assumed to be
\emph{sign-coherent}, that is, gathering either nonnegative,
nonpositive or
null parameters. To tackle this problem, we propose the \emph
{cooperative-Lasso} penalty.
We derive the optimality conditions defining the cooperative-Lasso estimate
for generalized linear models, and propose an efficient active set algorithm
suited to high-dimensional problems. We study the asymptotic consistency
of the estimator in the linear regression setup and derive its
irrepresentable conditions, which are milder
than the ones of the group-Lasso regarding the matching of groups with the
sparsity pattern of the true parameters.
We also address the problem of model selection in linear regression by
deriving an approximation of the degrees of freedom of the cooperative-Lasso
estimator. Simulations comparing the proposed estimator to the
group and sparse group-Lasso
comply with our theoretical results, showing consistent improvements
in support recovery for sign-coherent groups.
We finally propose two examples illustrating the wide applicability of the
cooperative-Lasso: first to the processing of ordinal variables, where the
penalty acts as a monotonicity prior; second to the processing of genomic
data, where the set of differentially expressed probes is enriched by
incorporating all the probes of the microarray that are related to the
corresponding genes.
\end{abstract}

%

\end{frontmatter}

\setcounter{footnote}{1}
\section{Introduction}

This paper addresses the problems of estimation and inference of parameters
when a group structure among parameters is known. We propose\vadjust{\goodbreak} a new
penalty for
the case where the groups are assumed to gather either nonpositive,
nonnegative or null parameters. All such groups will be referred to as
\textit{sign-coherent}.

As the main motivating example, we consider the linear regression model
%
\begin{equation}
\label{eqlinearreggroup}
Y = X \bbeta^\star+ \varepsilon
= \sum_{k=1}^K \sum_{j\in\group} X_j \beta_j^\star+ \varepsilon
,
\end{equation}
where $Y$ is a continuous response variable, $X=(X_1,\ldots,X_p)$ is a
vector of $p$
predictor variables, $\bbeta^\star$ is the vector of unknown
parameters and
$\varepsilon$ is a zero-mean Gaussian error variable with variance
$\sigma^2$.
The set of indexes $\{1,\ldots,p\}$ is partitioned into $K$ groups
$\{\group\}_{k=1}^K$ corresponding to predictors and parameters.
We will assume throughout this paper that $\bbeta^\star$ has few nonzero
coefficients, with sparsity and sign patterns governed by the groups~$\group$,
that is, groups being likely to gather either positive, negative or null
parameters.

The estimation and inference of $\bbeta^\star$ is based on training data,
consisting of a vector
$\mathbf{y}=(y_1,\ldots,y_n)^\intercal$ for responses and a
$n\times p$ design matrix $\mathbf{X}$ whose $j$th  column contains
$\mathbf{x}_j = (x_j^1,\ldots,x_j^n)^\intercal$, the $n$ observations
for variable $X_j$. For\vspace*{1pt} clarity, we assume that both $\mathbf{y}$
and $\{\mathbf{x}_j\}_{j=1,\ldots,p}$ are centered so as to eliminate the
intercept from fitting criteria.

Penalization methods that build on the $\ell_1$-norm, referred to as
\emph{Lasso} procedures (Least Absolute Shrinkage and Selection
Operator), are
now widely used to tackle simultaneously variable estimation and
selection in
sparse problems. Among these, the group-Lasso, independently proposed by
\citet{1998NIPSGrandvalet} and \citet{1999thesisBakin} and
later developed
by \citet{2006JRSSYuan}, uses the group
structure to define a shrinkage estimator of the form
%
\begin{equation}
\label{eqgrouplassolinear}
\hat{\bbeta}{}^{\mathrm{group}} = \argmin_{\bbeta\in\mathbb{R}^p}
 \Biggl\{
\frac{1}{2} \| \mathbf{y} - \mathbf{X}\bbeta \|^2 +
\lambda\sum_{k=1}^K w_k  \|\bbeta_{\mathcal{G}_k}  \|
 \Biggr\}
,
\end{equation}
where $\mathcal{G}_k$ is the subset of indices defining the $k$th
group of
variables and $\| \cdot \|$ is the Euclidean norm. The tuning
parameter $\lambda\geq 0$ controls the overall amount of penalty and
weights $w_k>0$ adapt the level of penalty within a given
group. Typically, one sets $w_k = \sqrt{p_k}$, where $p_k$ is the
cardinality of~$\group$ in order to adjust
shrinkage according to group sizes. The penalizer in~\eqref{eqgrouplassolinear} is known to induce sparsity at the
group level, setting a whole group of parameters to zero for values of
$\lambda$ which are large enough. Note that when we assign one group
to each predictor, we recover the original Lasso [\citet
{1996JRSSTibshirani}].

The algorithms for finding the group-Lasso estimator have considerably improved
recently.
\citet{2010preprintFoygel} develop a block-wise algorithm, where
each group of
coefficients is updated at a time, using a single line search that
provides the
exact optimal value for one group, considering all other coefficients fixed.
\citet{2008JRSSMeier} depart from linear regression in problem
\eqref{eqgrouplassolinear} by studying group-Lasso penalties for logistic
regression. Their block-coordinate descent method is applicable to
generalized linear models.
Here, we build on the subdifferential calculus approach originally
proposed by
\citet{2000JCGSOsborne} for the Lasso, whose active set algorithm
has been
adapted to the group-Lasso [\citet{2008ICMLRoth}].

Compared to the group-Lasso, this paper deals with a stronger
assumption regarding the group structure. Groups should not only reveal the
sparsity pattern, but they should also be relevant for sign patterns:
all coefficients within a group should be sign-coherent, that is, they should
either be null, nonpositive or nonnegative.
This desideratum arises often when the groups gather redundant or consonant
variables (a usual outcome when groups are defined from clusters of correlated
variables).
To perform this sign-coherent grouped variable selection, we propose a novel
penalty that we call the cooperative-Lasso, in short the \emph{coop-Lasso}.

The coop-Lasso is amenable to the selection of patterns
that cannot be achieved with the group-Lasso. This ability, which can be
observed for finite samples, also leads to consistency results under
the mildest
assumptions.
Indeed, the consistency results for the group-Lasso assume that the
set of nonzero coefficients of $\bbeta^\star$ is
an exact union of groups [\citet{2008JMLRBach}; \citet
{2008EJSNardi}], while
exact support recovery may be achieved with coop-Lasso when some zero
coefficients belong to a group having either positive or negative coefficients.
For example, with groups $\group[1]=\{1,2\}$ and $\group[2]=\{3,4,5\}
$, the
support of $\bbeta^\star=(-1,1,0,1,1)^\intercal$ may be recovered
with the
coop-Lasso, but not with the group-Lasso,
%
which may then deteriorate the performances of the Lasso [\citet
{2010preprintHuang}].
\citet{2010preprintFriedman} propose to overcome this restriction
by adding an
$\ell_1$ penalty to the objective function in~\eqref
{eqgrouplassolinear}, in
the vein of the hierarchical penalties of \citet{2009ASZhao}.
The new term provides additional flexibility but demands an additional tuning
parameter, while our approach takes a different stance by assuming
sign-coherence, with the benefit of requiring a single tuning parameter.

Section~\ref{secdata} describes two applications where sign-coherence
is a
sensible assumption.
The first one considers ordered categorical data, which are common in
regression and classification. The coop-Lasso can then be used to
induce a monotonic
response to the ordered levels of a covariate, without translating each
level of
the categorical variable into a prescribed quantitative value.
The second application describes the situation where redundancy in measurements
causes sign-coherence to be expected.
Similar behaviors should be observed when features have been grouped by
a clustering
algorithm such as average linkage hierarchical clustering, which are nowadays
routinely used for grouping genes in microarray data analysis
[\citet{1998PNASEisen}; \citet{2006BSPark}; \citet
{2007BMCMa}].\vadjust{\goodbreak}

Finally, in numerous problems of multiple inference, the sign-coherence
assumption is also reasonable: when predicting closely related
responses (e.g.,
regressing male and female life expectancy against economic and social
variables) or when analyzing multilevel data (e.g., predicting academic
achievement against individual factors across schools), the set of coefficients
associated to a predictor (resp., for all response variables or
all data
clusters) forms a group that can often be considered as sign-coherent because
effects can be assumed to be qualitatively similar.
Along these lines, we successfully applied the coop-Lasso penalizer for the
joint inference of several network structures [\citet{2010SCChiquet}].


The rest of the paper is organized as follows: Section~\ref{seccoop}
presents the coop-Lasso penalty, with
the derivation of the optimality conditions which are the basis
for an active set algorithm.
Consistency results and the
associated irrepresentable conditions are given in
Section~\ref{secconsistency}. In
Section~\ref{secmodelselection} we
derive an approximation of the degrees of freedom that
can be used in the Bayesian Information Criterion (BIC) and the Akaike
Information Criterion (AIC)
for model selection. Section~\ref{secnumerical} is dedicated
to simulations assessing the performances of the coop-Lasso in terms of sparsity
pattern recovery, parameters estimation and robustness.
Section~\ref{secdata} considers real data sets, with ordinal and
continuous covariates.
Note that all proofs are postponed until the \hyperref[appm]{Appendix}.


\section{Cooperative-Lasso}\label{seccoop}

\subsection{Definitions and optimality conditions}

\textit{Group-norm and coop-norm.}
We define a group structure by setting a partition of the index set
$\mathcal{I}=\{1,\ldots,p\}$, that is,
\[
\mathcal{I}=\bigcup_{k=1}^K\group   \qquad \mbox{with }
\group\cap\group[\ell]=\varnothing
\mbox{ for } k\neq\ell.
\]
Let $\mathbf{v} = (v_1,\ldots,v_p)^\intercal \in \Rset^p$ and
$p_k
$
denote the cardinality of group $k$. We define
$\mathbf{v}_{\group} \in \Rset^{p_k}$ as the vector $(v_j)_{j\in
\group}$.
For the chosen groups $\{\group\}_{k=1}^K$, the group-Lasso norm reads
%
\begin{equation}
\label{eqgroupnorm}
\| \mathbf{v} \|_{\mathrm{group}} = \sum_{k=1}^K
w_k  \| \mathbf{v}_{\group} \|
,
\end{equation}
where $w_k>0$ are fixed parameters enabling to adapt the amount of
penalty for each group. Likewise, the sparse group-Lasso norm
[\citet{2010preprintFriedman}] is defined as a convex combination
of the
group-Lasso and the $\ell_1$ norms:
%
\begin{equation}
\label{eqsparsegroupnorm}
\| \mathbf{v} \|_{\mathrm{sgl}} = \alpha\| \mathbf{v} \|_{\mathrm{group}}
+ (1-\alpha) \| \mathbf{v} \|_{\mathrm{1}}
,
\end{equation}
where $\alpha$ is meant to be a tuning parameter, but may be fixed to $1/2$
[\citet{2010preprintFriedman}; \citet{Zhou01102010}]. We
will always set it to this
default value in what follows.

Let $\mathbf{v}^+ = (v^+_1,\ldots,v^+_p)^\intercal$ and
$\mathbf{v}^- = (v^-_1,\ldots,v^-_p)^\intercal $ be the componentwise
positive and negative part of $\mathbf{v}$, that is, $ v^+_j =
\max(0,v_j)$ and $v^-_j = \max(0,-v_j)$, respectively. We call
\emph{coop-norm} of $\mathbf{v}$ the sum of group-norms on
$\mathbf{v}^+$ and $\mathbf{v}^-$,
\[
 \|\mathbf{v} \|_{\mathrm{coop}} =
 \|\mathbf{v}^+ \|_{\mathrm{group}} +
 \|\mathbf{v}^- \|_{\mathrm{group}}=
\sum_{k=1}^K w_k  ( \|\mathbf{v}_{\group}^+ \| +
 \|\mathbf{v}_{\group}^- \| )
,
\]
which is clearly a norm on $\mathbb{R}^p$.

The coop-Lasso
estimate of $\bbeta^\star$ as defined in \eqref{eqlinearreggroup}
is
%
\begin{equation}
\label{eqcooplassolinear}
\hatbbetacoop = \argmin_{\bbeta\in\mathbb{R}^p}
L(\bbeta) \qquad\mbox{with } L(\bbeta) =
\frac{1}{2} \|\mathbf{y}- \mathbf{X}\bbeta \|^2
+ \lambda
\| \bbeta\|_{\mathrm{coop}}
,
\end{equation}
%
where $\lambda \geq 0$ is a tuning parameter common to all
groups. Appropriate choices for $\lambda$ 
will be discussed in Sections \ref{secmodelselection} and \ref{secconsistency}
dealing with model selection and consistency, respectively.

\begin{figure}[t!]

\includegraphics{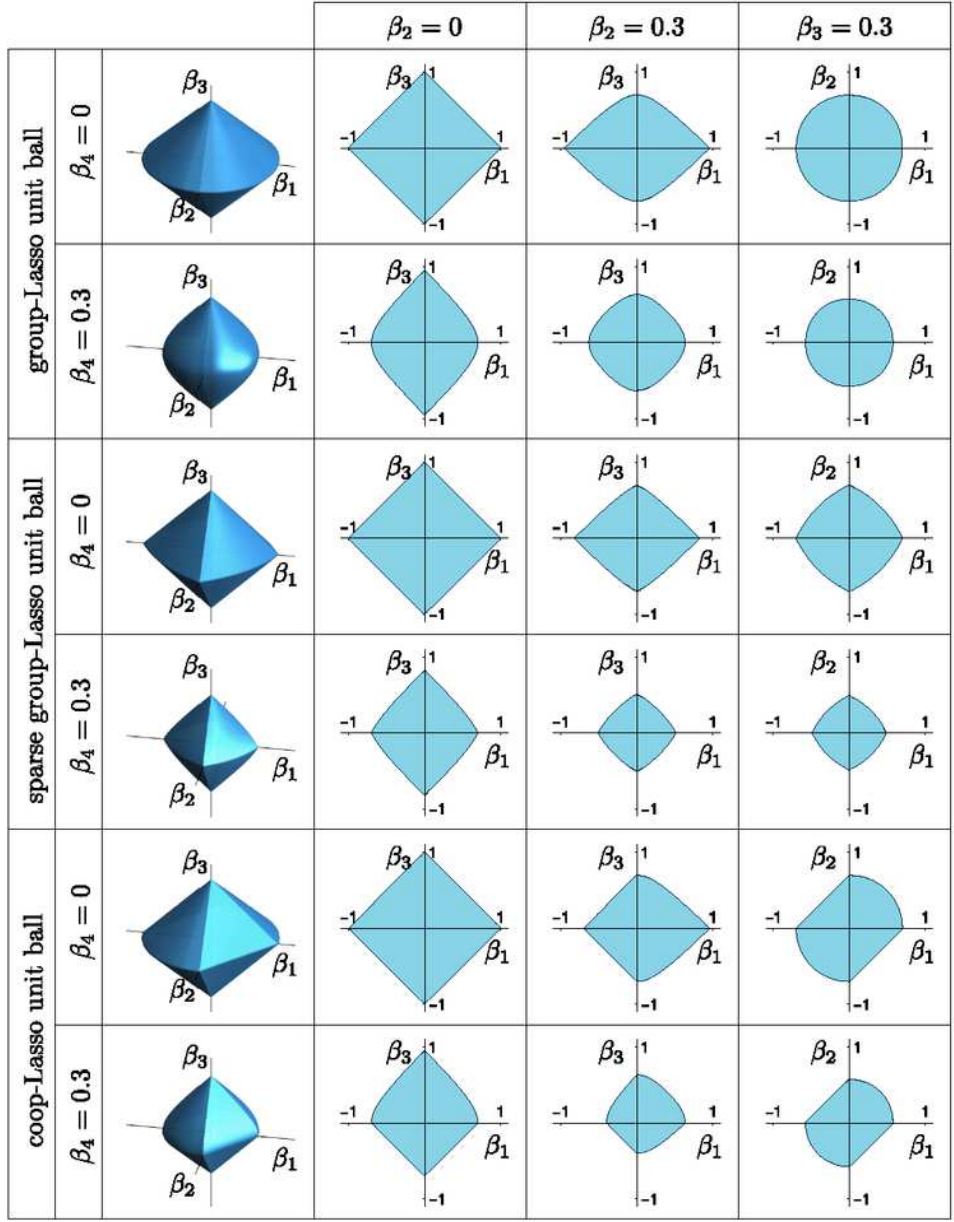}

\caption{Feasible sets for the coop-Lasso, group-Lasso and sparse
group-Lasso penalties.
First column: cuts through $(\beta_{1},\beta_{2},\beta_{3})$ at
$\beta_{4}=0$ and $\beta_{4}=0.3$:
$(\beta_{1},\beta_{2})$ span the horizontal plane and $\beta_{3}$ is
on the vertical axis; second and third columns: cuts through
$(\beta_{1},\beta_{3})$ at various values of $(\beta_{2},\beta_{4})$;
last column: cuts through $(\beta_{1},\beta_{2})$ at various values
of $(\beta_{3},\beta_{4})$.}
\label{figcoop-norm}
\end{figure}

Illustrations of the group, sparse group and coop norms are given in
Figure~\ref{figcoop-norm} for a vector
 $\bbeta= (\beta_{1},\beta_{2},\beta_{3},\beta
_{4})^\intercal$
with two groups $\group[1] = \{1,2\}$ and $\group[2] = \{3,4\}$.
We represent several views of the unit ball for each of these norms.
For the coop-norm, this ball represents the set of feasible
solutions for an optimization problem equivalent to~\eqref{eqcooplassolinear},
where the sum of squared residuals is minimized under unitary
constraints on
$\| \bbeta\|_{\mathrm{coop}}$. The same interpretation holds for the
group and
sparse group norms, provided the sum of squared residuals is minimized
under unitary constraints on $\| \bbeta\|_{\mathrm{group}}$ and $\|
\bbeta\|_{\mathrm{sgl}}$, respectively.

%

These plots provide some insight into the sparsity pattern that
originates from
the penalties, since sparsity is related to the singularities of the
boundary of
the feasible set.
First, consider the group-Lasso: the first row illustrates that when
$\beta_{4}$
is null its group companion $\beta_{3}$ may also be exactly zero
(corners on the
boundary at $\beta_{3}=0$), while the second row shows that this event is
improbable when $\beta_{4}$ differs from zero (smooth boundary at
$\beta_{3}=0$).
The second and third columns display the same type of relationships within
$\group[1]$ between $\beta_{2}$ and $\beta_{1}$, which are expected
due to the
symmetries of the unit ball.
The last column displays $\ell_2$ balls, which characterize the within-groups
feasibility subsets, showing that once a group is activated, all its members
will be nonzero.

Now, consider the sparse group-norm: the combination of the group and Lasso
penalties has uniformly shrunk the feasible set toward the Lasso $\ell
_1$ unit
ball, thus creating new edges that provide a chance to zero any
parameter in any
situation, with an elastic-net-like penalty [\citet{2005JRSSZou}]
within and
between groups. The comparison of the last two columns illustrates that the
differentiation between the within-group and between group penalties is less
marked than for the group-Lasso.

Finally, consider the coop-norm: compared to the group-norm,
there are also additional discontinuities resulting in new edges on the
3-D plots.
While the sparse group-Lasso edges where created by a uniform shrinking
toward the $\ell_1$ unit ball, the coop-Lasso new edges result from slicing
the group-Lasso unit ball, depriving sign-incoherent orthants from some
of the
group-Lasso feasible solutions ($\|\bbeta\|_{\mathrm{coop}} > \| \bbeta
\|_{\mathrm{group}}$ in these regions).
Note that, in general, there are less new edges than with the sparse
group-Lasso, since the new opportunities to zero some coefficients are limited
to the case where the group-Lasso would have allowed a solution with opposite
signs within a group.
The crucial difference with the group and sparse group-Lasso is the
loss of the axial
symmetry when some variables are nonzero:
decoupling the positive and negative parts of the regression
coefficients favors
solutions where signs match within a group.
Slicing of the unit group-norm ball does not affect the positive
and negative orthants, but large areas corresponding to sign mismatches have
been peeled off, as best seen on the last column, which also
illustrates the
strong differentiation between within-group and between-group penalties.

Before stating the optimality conditions for
problem~\eqref{eqcooplassolinear}, we introduce some notation related
to the
sparsity pattern of parameters, which will be required to express the necessary
and sufficient condition for optimality.
First, we recall that the unknown vector of parameters $\bbeta^\star$
is typically sparse; its support is denoted $\supp=\{j, \bbeta
_j^\star\neq0\}$
and $\supp^c=\{j, \bbeta_j^\star= 0\}$ is the complementary set of
true zeros.
Once the problem has been supplied with a group structure, we define
$\supp_k =
\supp\cap\group$ and $\supp_k^c = \supp^c \cap\group$ as the
sets of
relevant, respectively irrelevant, predictors within group $k$, for all
$k=1,\ldots,K$. Similar notation $\supp(\bbeta)$, $\supp_k(\bbeta)$ and
$\supp_k^c(\bbeta)$ is defined for an arbitrary vector $\bbeta\in
\Rset^p$.
Furthermore, for clarity and brevity, we introduce the functions
$\{\bvarphi_j\}_{j=1}^{p}$,
which return the componentwise positive or negative part of a vector according
to the sign of its $j$th element
, that is,
$\forall k \in\{1,\ldots,K\} , \forall j\in\group , \forall
\mathbf{v}\in\Rset^{p_k},$
%
\begin{equation}
\label{eqphij}
\bvarphi_j(\mathbf{v}) = (\sign(v_j)\mathbf{v})^+ =
\cases{\displaystyle
\mathbf{0} ,&\quad   if   $v_j = 0$,\cr\displaystyle
\mathbf{v}^+ ,&\quad   if   $v_j > 0$,\cr\displaystyle
\mathbf{v}^- ,&\quad   if   $v_j < 0$.
}
\end{equation}

\textit{Optimality conditions.}
The objective function $L$ in \eqref{eqcooplassolinear} is
continuous and coercive, thus problem~\eqref{eqcooplassolinear}
admits at least one minimum.
If $\mathbf{X}$ has rank~$p$, then the minimum is unique since~$L$ is strictly
convex.
Furthermore,~$L$ is smooth, except at some locations
with zero coefficients, due to the singularities of the coop-norm.
Since $L$ is convex, a necessary and sufficient condition for the
optimality of $\bbeta$ is that the null vector $\mathbf{0}$ belongs to
the subdifferential of $L$ whose expression is provided in the
following lemma.


\begin{lemma}\label{lemsubdifferential}
For all $\bbeta\in\Rset^p$, the subdifferential of the objective
function of
problem~\eqref{eqcooplassolinear} is
%
\begin{equation}
\partial_{\bbeta} L(\bbeta) =
 \{ \mathbf{v} \in\Rset^p \dvtx
\mathbf{v} = \mathbf{X}^\intercal
(\mathbf{X}\bbeta-\mathbf{y}) + \lambda
\btheta \},
\end{equation}
where $\btheta\in\Rset^p$ is any vector belonging to the
subdifferential of
the coop-norm, that is,
%
\begin{subequation}
\label{eqsubgradientcompact}
%
\begin{eqnarray}\label{eqsubgradientcompactb}
\forall k &\in&\{1,\ldots,K\} , \forall j\in\supp_k(\bbeta)   \qquad  \theta_j = \frac{w_k\beta_j} { \|
\bvarphi_j(\bbeta_{\group})  \|} ,
\\\label{eqsubgradientcompacta}
\forall k &\in&\{1,\ldots,K\} , \forall j\in\supp_k^c(\bbeta)   \qquad   \| \bvarphi_j (\btheta_{\group}  ) \|
\leq w_k .
\end{eqnarray}
\end{subequation}
\end{lemma}

The following optimality conditions, which result directly from
Lemma~\ref{lemsubdifferential}, are an essential building block of the
algorithm we propose to compute the coop-Lasso estimate. They also
provide an
important basis for showing the consistency results.

%
\begin{theorem}\label{thmoptimality}
Problem \eqref{eqcooplassolinear} admits at least one solution, which is
unique if  $\mathbf{X}$ has rank $p$.
All critical points $\bbeta$ of the objective function $L$ verifying the
following conditions are global minima:
%
\begin{subequation}
\label{eqoptimalityall}
%
\begin{eqnarray}
\label{eqoptimality}
\forall k &\in&\{1,\ldots,K\} , \forall j\in\supp_k(\bbeta)  \qquad \mathbf{x}^\intercal_j(\mathbf{X}\bbeta-\mathbf{y}) +
\frac{\lambda w_k\beta_j}{\|\bvarphi_j(\bbeta_{\group})\|} = 0
, \\
\label{eqoptimalityzero}
\forall k &\in&\{1,\ldots,K\} , \forall j\in\supp_k^c(\bbeta)  \qquad  \bigl\|
\bvarphi_j \bigl((\mathbf{X}_{\centerdot\group})^{\intercal
}(\mathbf{X}\bbeta-\mathbf{y}) \bigr)
 \bigr\| \leq\lambda w_k
,
\end{eqnarray}
\end{subequation}
where $\mathbf{X}_{\centerdot\group}$ is the submatrix of $\mathbf
{X}$ with
all rows and columns indexed by $\group $.
\end{theorem}

Note here an important distinction compared to the group-Lasso, where
the optimality
conditions are expressed solely according to the groups $\group$
[see, e.g., \citet{2008ICMLRoth}].
Hence, while the sparsity pattern of the solution is strongly
constrained by the
predefined group structure in the group-Lasso, deviations from this structure
are possible for the coop-Lasso.
The asymptotic analysis of Section~\ref{secconsistency} confirms that exact
support recovery is possible even when the support of $\bbeta^\star$
cannot be
expressed as a simple union of groups, provided the groups intersecting
the true
support are sign-coherent.

\subsection{Algorithm}

The efficient approaches developed for the Lasso take advantage of the sparsity
of the solution by solving a series of small linear systems, whose
sizes are
incrementally increased/decreased [\citet{2000JCGSOsborne}]. This
approach was
pursued for the group-Lasso [\citet{2008ICMLRoth}] and we
proposed an
algorithm in the same vein for the coop-Lasso in the framework of
multiple network inference [\citet{2010SCChiquet}].
We provide here a more detailed description of the latter in the
specific context
of linear regression.

The algorithm starts
from a sparse initial guess, say, $\bbeta=0$, and iterates two steps:\vadjust{\goodbreak}
%
\renewcommand\thelonglist{\arabic{longlist}}
\renewcommand\labellonglist{\arabic{longlist}.}
\begin{longlist}[2.]
\item\label{itemalgostep1} The first step solves problem~\eqref
{eqcooplassolinear} with respect to
$\bbeta_{ \mathcal{A}}$, the subset of ``active'' variables, currently
identified as being nonzero.
At this stage the current feasible set is restricted to the orthants
where the
gradient of the coop-norm has no discontinuities: the optimization
problem is
thus smooth.
One or more variables may then be declared inactive if the current optimal~$\bbeta_{ \mathcal{A}}$ reaches the boundary of the current feasible set.
\item\label{itemalgostep2} The second step assesses the completeness
of the set $\mathcal{A}$,
by checking the optimality conditions with respect to inactive
variables. We add a group
that violates these conditions.
In our implementation, we pick the one that most violates the optimality
condition, since this strategy has been observed to require few changes
in the active set.
When no such violation exists, the current solution is optimal.
\end{longlist}

\begin{algorithm}[t]
  \hspace*{22pt}\begin{minipage}{327pt}
\nlset{{\normalsize Init.}} Start from a feasible $\bbeta\leftarrow\bbeta^0$
\begin{eqnarray*}
\mathcal{A}_{+} &\leftarrow&\{j\in\group\dvtx \|\bbeta_{\group}^+\| >
0  ,  k=1,\ldots,K\}, \\
\mathcal{A}_{-} &\leftarrow&\{j\in\group\dvtx \|\bbeta_{\group}^-\| >
0  ,  k=1,\ldots,K\}.
\end{eqnarray*}
\nlset{{\normalsize Step \ref{itemalgostep1}}}On $\mathcal{A} \leftarrow\mathcal
{A}_{+} \cup\mathcal{A}_{-}$, find
a solution to the smooth problem
%
\begin{eqnarray}
\bbeta_{ \mathcal{A}} \leftarrow
\argmin_{\mathbf{v}\in\mathbb{R}^{|\mathcal{A}|}} \frac
{1}{2} \|\mathbf{y}-\mathbf{X}_{\centerdot\mathcal{A}} \mathbf
{v}  \|^2
+ \lambda \| \mathbf{v} \|_{\mathrm{coop}}\nonumber\\
\eqntext{\mbox{s.t. }
\cases{\displaystyle
v_j \geq0 ,&\quad   if   $j\in\mathcal{A}_{+} \cap\mathcal{A}_{-}^c
$,\cr\displaystyle
v_j \leq0 ,&\quad   if  $ j\in\mathcal{A}_{-} \cap\mathcal{A}_{+}^c
$,
}}
\end{eqnarray}
where $\mathcal{A}_{-}^c$ and $\mathcal{A}_{+}^c$ are the
complementary sets
of $\mathcal{A}_{-}$ and $\mathcal{A}_{+}$, respectively.

Identify groups inactivated during optimization
\begin{eqnarray*}
\mathcal{A}_{+} &\leftarrow&\mathcal{A}_{+}\bigm\backslash\Bigl\{
j\in\group\subseteq\mathcal{A}_{+}\dvtx \|\bbeta_{\group}^+\| = 0
\\
&&\hphantom{\mathcal{A}_{+}\setminus\Bigl\{} \mbox{and } \min_{\mathbf{v} \in\partial_{\bbeta_{\group}}
L(\bbeta)}  \|\mathbf{v^-} \|=0  ,  k=1,\ldots,K\Bigr\} ,\\
\mathcal{A}_{-} &\leftarrow&\mathcal{A}_{-}\bigm\backslash\Bigl\{
j\in\group\subseteq\mathcal{A}_{-}\dvtx \|\bbeta_{\group}^-\| = 0
\\
&&\hphantom{\mathcal{A}_{-}\bigm\backslash\Bigl\{}\mbox{and } \min_{\mathbf{v} \in\partial_{\bbeta_{\group}}
L(\bbeta)}  \|\mathbf{v^+} \|=0  ,  k=1,\ldots,K\Bigr\}.
\end{eqnarray*}
\nlset{{\normalsize Step \ref{itemalgostep2}}}
Identify the greatest violation of optimality conditions:
\begin{eqnarray*}
 g^k_+ &\leftarrow&\min_{\mathbf{v} \in\partial
_{\bbeta_{\group}}  L(\bbeta)}  \|\mathbf{v^+} \|
, \qquad
 q \leftarrow\argmax_{k} g^k_+ , \\
 g^k_- &\leftarrow&\min_{\mathbf{v} \in\partial
_{\bbeta_{\group}}  L(\bbeta)}  \|\mathbf{v^-} \|
, \qquad
 r \leftarrow\argmax_{k} g^k_-
\end{eqnarray*}
%
\eIf{$\max(g^q_+,g^r_-)=0$}{
Stop and return $\bbeta$, which is
optimal
}{
\lIf{$g^q_+ > g^r_-$}{ $\mathcal{A}_{-} \leftarrow\mathcal{A}_{-}
\cup
\group[q]$ }
\lElse{ $\mathcal{A}_{+} \leftarrow\mathcal{A}_{+} \cup\group[r]$  }\\
Repeat Steps 1 and 2 until convergence  }
\vspace*{1pt}
\end{minipage}
\caption{Coop-Lasso fitting algorithm}
\label{algoactiveconstraint}
\end{algorithm}

These two steps outline the algorithm, which is detailed in more
technical terms in Algorithm~\ref{algoactiveconstraint}. The
principle is readily applied to any generalized linear model by simply
defining the appropriate objective function $L$. In our current
implementation (a pre-release of our \texttt{R}-package \texttt{scoop}
is available at \url{http://stat.genopole.cnrs.fr/logiciels/scoop})
the linear and logistic regression models are implemented using either
Broyden--Fletcher--Goldfarb--Shanno (BFGS) quasi-Newton updates with
box constraints, or proximal methods [\citet{2009SIAMBeck}] to solve
the smooth optimization problem in Step~\ref{itemalgostep1}.

Finally, note that to compute a series of solutions along the
regularization path for
problem~\eqref{eqcooplassolinear}, we simply choose a series of penalties
$\lambda^1=\lambda_{\mathrm{max}} >\cdots > \lambda^l>\cdots > \lambda
^L =
\lambda_{\mathrm{min}} \geq0$ such that $\hatbbetacoop(\lambda
_{\mathrm{max}})=\mathbf{0}$, that is,
\[
\lambda_{\mathrm{max}} = \max_{k\in\{1,\ldots,K\}} \max_{j\in
\group} \frac{1}{w_k}
 \|\bvarphi_j ((\mathbf{X}_{\centerdot\group})^\intercal
\mathbf{y} ) \|
.
\]
We then use the usual warm start strategy, where the feasible initial
guess for
$\hatbbetacoop(\lambda^{l})$, the coop-Lasso estimate with penalty parameter
$\lambda^l$, is initialized with $\hatbbetacoop(\lambda^{l-1})$.

\subsection{Orthonormal design case}

The orthonormal design case, where\break $\mathbf{X}^\intercal\mathbf{X} =
\mathbf{I}_p$, has been providing useful insights for penalization techniques
regarding the effects of shrinkage.
Indeed, in this particular case, most usual shrinkage estimators can be
expressed in closed-form as functions of the ordinary least squares (OLS)
estimate.
These expressions pave the way for the derivation of approximations of the
degrees of freedom
[\citet{1996JRSSTibshirani}; \citet{2006JRSSYuan} and
Section~\ref{secmodelselection}], which
may be convenient for model selection in the absence of exact formulae.

In the orthonormal setting, for any $\beta_j$, we have
$\mathbf{x}_j^\intercal(\mathbf{X}\bbeta-\mathbf{y}) = \beta_j -
\hatbeta_j^{\mathrm{ols}}$.
The optimality conditions~\eqref{eqoptimality} and~\eqref{eqoptimalityzero}
can then be written as
%
%
\begin{equation}
\label{eqcooportho}
\forall k \in\{1,\ldots,K\} , \forall j\in\group  \qquad \hatbeta_j^{\mathrm{coop}} =  \biggl(
1-\frac{\lambda w_k}{ \|\bvarphi_j(\hatbbetaols_{\group}) \|}
 \biggr)^{  +}
\hatbeta_j^{\mathrm{ols}}
.
\end{equation}
For reference, we recall the solution to the group-Lasso [\citet
{2006JRSSYuan}] in the
same condition
%
\begin{equation}
\label{eqgrouportho}
\forall k \in\{1,\ldots,K\} , \forall j\in\group  \qquad
\hatbeta_j^{\mathrm{group}} = \biggl (
1-\frac{\lambda w_k}{ \| \hatbbetaols_{\group}  \|}
 \biggr)^{  +}
\hatbeta_j^{\mathrm{ols}}
,
\end{equation}
while the Lasso solution [\citet{1996JRSSTibshirani}] is
%
\begin{equation}
\label{eqlassoortho}
\forall j\in\{1,\ldots,p\}  \qquad
\hatbeta_j^{\mathrm{lasso}} =  \biggl(
1-\frac{\lambda}{ | \hatbeta_j^{\mathrm{ols}}  |}
 \biggr)^{  +}
\hatbeta_j^{\mathrm{ols}}
.
\end{equation}

Equations~\eqref{eqcooportho}--\eqref{eqlassoortho} reveal strong
commonalities.
First, the coefficients of these shrinkage estimators are of the sign
of the OLS estimates.
Second, the norm used in the penalty defines a region where small OLS
coefficients are shrunk to zero, while large ones are shrunk inversely
proportional to this norm.
Finally, by grouping the terms corresponding to one
group in equations~\eqref{eqcooportho}--\eqref{eqgrouportho}, a uniform
translation effect, analogous to the one observed for the Lasso, comes into
view:
%
\begin{eqnarray}
\label{eqcoopnormortho}
&\displaystyle \forall k  \in \{1,\ldots,K\} , \forall j\in\group  \qquad
 \|\bvarphi_j(\hatbbetacoop_{\group})  \| =  \bigl(
 \| \bvarphi_j(\hatbbetaols_{\group}) \| - \lambda w_k
 \bigr)^{  +}
,& \nonumber\\
&\displaystyle \forall k  \in \{1,\ldots,K\}  \qquad
 \| \hatbbetagroup_{\group}  \| =  (
 \| \hatbbetaols_{\group}  \|- \lambda w_k
 )^{  +}
,&\\
&\displaystyle \forall j \in \{1,\ldots,p\}  \qquad
 | \hatbeta_j^{\mathrm{lasso}}  | =  (
 | \hatbeta_j^{\mathrm{ols}}  | - \lambda w_k
 )^{  +}
.&\nonumber
\end{eqnarray}
The group-Lasso \eqref{eqgrouportho} differs primarily
from the Lasso \eqref{eqlassoortho} owing to the common penalty
${\lambda
w_k}/{\| \hatbbetaols_{\group}\|}$ for all the coefficients belonging
to group
$k$. The magnitude of shrinkage is determined by all within-group OLS
coefficients,
and is thus radically different from a ridge regression penalty in this regard.
For the coop-Lasso estimator \eqref{eqcooportho}, two penalties
possibly apply
to group $k$, for the positive and the negative OLS coefficients, respectively.
If all within-group OLS coefficients are of the same sign, coop-Lasso
is identical
to group-Lasso; if some signs disagree, the magnitude of the penalty
only depends on the within-group OLS coefficients with an identical
sign. In the
extreme case where exactly one OLS coefficient is positive/negative, the
coop-penalty is identical to a Lasso penalty on this coefficient.

Note that such a simple analytical formulation is not available for the sparse
group-Lasso estimate $\hatbbeta{}^{\mathrm{sgl}}$, but an expression can
be obtained
by chaining two simple shrinkage operations.
Introducing an intermediate solution $\tilde{\bbeta}{}^\mathrm{sgl}$, we have,
$\forall k \in\{1,\ldots,K\}$ and $\forall j\in\group ,$
%
%
\begin{equation}
\label{eqsparsegrouportho}\qquad
\hatbeta_j^{\mathrm{sgl}} =  \biggl(
1-\frac{\lambda(1-\alpha) w_k}{ \|\tilde{\bbeta}{}^\mathrm
{sgl}_{\group} \|}
 \biggr)^{  +}
\tilde{\beta}{}^\mathrm{sgl}_{j}
 \qquad  \mbox{where }
\tilde{\beta}{}^\mathrm{sgl}_{j} =  \biggl(
1-\frac{\lambda\alpha}{ | \hatbeta_j^{\mathrm{ols}}  |}
 \biggr)^{  +}
\hatbeta_j^{\mathrm{ols}}
.
\end{equation}
The intermediate solution $\tilde{\bbeta}{}^\mathrm{sgl}$ is the Lasso estimator
with penalty parameter~$\lambda\alpha$, which acts as the OLS
estimate for a
group-Lasso of parameter $\lambda(1-\alpha)$.

\begin{figure}

\includegraphics{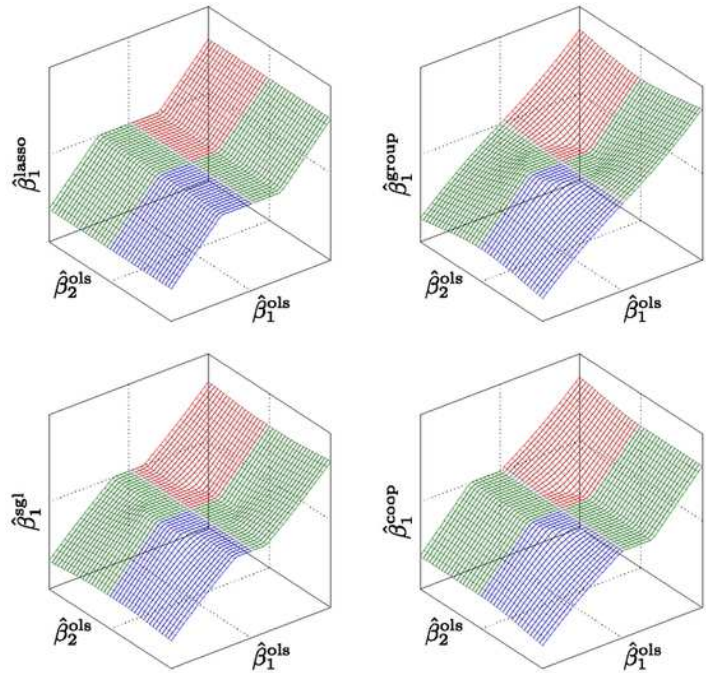}

\caption{Lasso, group, sparse group and coop Lasso
coefficient estimates, for a group with 2 elements $\group=\{1,2\}$,
as a
function of the OLS coefficients. The colors emphasize the positive and
negative quadrants of the
$(\hatbeta{}^{\mathrm{ols}}_1,\hatbeta{}^{\mathrm{ols}}_2)$ plane, with red and
blue,
respectively.}\label{figorthonormal}\vspace*{-3pt}
\end{figure}

Figure~\ref{figorthonormal} provides a visual representation of equations
\eqref{eqcooportho}--\eqref{eqlassoortho} and \eqref{eqsparsegrouportho}
for a group with two components, say, $\group=\{1,2\}$.
%
%
We plot $\hatbeta_1^{\mathrm{lasso}},
\hatbeta_1^{\mathrm{group}}, \hatbeta_1^{\mathrm{sgl}}$ and $\hatbeta
_1^{\mathrm{coop}}$ as functions of
$(\hatbeta_1^{\mathrm{ols}},\hatbeta_2^{\mathrm{ols}})$.
Top-left, the Lasso translates the $\hatbeta_1^{\mathrm{ols}}$
coefficient toward zero, eventually truncating them at zero, regardless of~$\hatbeta_2^{\mathrm{ols}}$: there is no interaction between coefficients.
The group-Lasso, top-right, has a nonlinear shrinking behavior (quite
different from the Lasso or ridge penalties in this respect) and sets
$\hatbeta_1^{\mathrm{group}}$ to zero within a Euclidean ball centered
at zero.
The sparse group-Lasso, bottom-left, is a hybrid of Lasso and
group-Lasso, whose shrinking behavior lies between its two ancestors.
Bottom-right, the coop-Lasso appears as another form of cross-breed, identical
to the group-Lasso in the positive and negative quadrants, and
identical to the
Lasso when the signs of the OLS coefficients mismatch.
For groups with more than two components, intermediate solutions would be
possible.
This behavior is shown to allow for some flexibility
with respect to the predefined group structure in the following consistency
analysis.

\section{Consistency}\label{secconsistency}

Beyond its sanity-check value, a consistency analysis brings along an
appreciation of the strengths and limitations of an estimation scheme.
Here we concentrate on the estimation of the support of the parameter
vector, that is, the position of its zero entries.
Our proof technique is drawn from the previous works on the Lasso
[\citet{2007JRSSYuan}] and the group-Lasso
[\citet{2008JMLRBach}].\vadjust{\goodbreak}

In this type of analysis, some assumptions on the joint distribution
of $(X,Y)$ are required to guarantee the convergence of empirical covariances.
For the sake of simplicity and coherence, we keep assuming that data are
centered so that we have zero mean random variables and
$\bPsi= \mathbb{E} [XX^\intercal ]$ is the covariance
matrix of $X$:
\begin{longlist}[(A2)]
\item[(A1)] $X$ and $Y$ have finite 4th order
moments $\mathbb{E} [\|X\|^4 ] < \infty$, $\mathbb
{E} [Y^4 ] < \infty$.
\item[(A2)] The covariance matrix $\bPsi= \mathbb{E}
[XX^\intercal ] \in\Rset^{p \times p}$ is invertible.
\end{longlist}

In addition to these standard technical assumptions, we need a more specific
one, substantially avoiding situations where the coop-Lasso will almost never
recover the true support:
%
\begin{longlist}[(A3)]
\item[(A3)]
All sign-incoherent groups are included in the true
support:
$\forall k\in\{1,\ldots,K\}$, if
$\|(\bbeta_{\group}^\star)^+\| >0$ and
$\|(\bbeta_{\group}^\star)^-\| >0$, then $\forall j \in\group$,
$\bbeta^\star_j \neq0$.
\end{longlist}
Note that this latter assumption is less stringent than the one
required for the
group-Lasso since it does not require that each group of variables
should either
be included in or excluded from the support.
For the coop-Lasso, sign-coherent groups may intersect the support.

The spurious relationships that may arise from confounding variables are
controlled by the so-called strong irrepresentable condition, which guarantees
support recovery for the Lasso [\citet{2007JRSSYuan}] and the group-Lasso
[\citet{2008JMLRBach}].
We now introduce suitable variants of these conditions for the coop-Lasso.
They result in two assumptions: a general one, on the
magnitude of correlations between relevant and irrelevant variables,
and a~more
specific one for groups which intersect the support, on the sign of
correlations.
These conditions will be expressed in a compact vectorial form using the
diagonal weighting matrix $\mathbf{D}(\bbeta)$ such that,
%
\begin{equation}\label{eqdefD}
\forall k \in\{1,\ldots,K\}, \forall j\in\supp_k(\bbeta)  \qquad
 (\mathbf{D}(\bbeta) )_{jj} = w_k \|\bvarphi_j(\bbeta
_{\group})\|^{-1}
.
\end{equation}\vspace*{-\baselineskip}
%
\begin{longlist}[(A4)]
\item[(A4)] For every group $\group$ including at least one null
coefficient (i.e., such that $\beta^\star_{j} = 0$ for some $j \in\group$ or, equivalently,
$\supp^c_k \neq\varnothing$), there exists $\eta>0$ such that
%
\begin{equation}
  \frac{1}{w_k} \max( \| (\bPsi_{\supp_k^c\supp} \bPsi_{\supp\supp}^{-1}
\mathbf{D}(\bbeta^\star_{\supp})\bbeta^\star_{\supp})^+ \|,
\| (\bPsi_{\supp_k^c \supp} \bPsi_{\supp\supp}^{-1}
\mathbf{D}(\bbeta^\star_{\supp})\bbeta^\star_{\supp})^-
\| ) \leq1-\eta
 , \label{thICnoPnoN}\hspace*{-35pt}
\end{equation}
where $\bPsi_{\cal{ST}}$ is the submatrix of $\bPsi$ with lines and columns
respectively indexed by
$\cal{S}$ and $\cal{T}$.
\end{longlist}
\begin{longlist}[(A5)]
\item[(A5)] For every group $\group$ intersecting the support and including
either positive or negative coefficients, letting $\nu_k$ be the sign
of these
coefficients [$\nu_k = 1$ if $\|(\bbeta_{\group}^\star)^+\| >0$ and
$\nu_k=-1$
if $\|(\bbeta_{\group}^\star)^-\| >0$], the following inequalities should
hold:
%
\begin{equation}
\nu_k \bPsi_{\supp^c_k \supp} \bPsi_{\supp\supp}^{-1}
\mathbf{D}(\bbeta^\star_{\supp})\bbeta^\star_{\supp}
\preceq\mathbf{0}
,
\label{thICPorN}
\end{equation}
where $\preceq$ denotes componentwise inequality.\vadjust{\goodbreak}
\end{longlist}
Note that the irrepresentable condition for the group-Lasso only considers
correlations between groups included and excluded from the support.
It is otherwise similar to \eqref{thICnoPnoN}, except that the
elements of the
weighting matrix~$\mathbf D$ are $w_k \|\bbeta_{\group}\|^{-1}$ and
that the
$\ell_2$ norm replaces $ \max( \| (\cdot)^+ \|, \| (\cdot)^-\| ) $.

We now have all the components for stating the coop-Lasso consistency theorem,
which will consider the following normalized (equivalent) form of the
optimization problem \eqref{eqcooplassolinear} to allow a direct comparison
with the known similar results previously stated for the Lasso and group-Lasso
[\citet{2007JRSSYuan}; \citet{2008JMLRBach}]:
%
\begin{equation}
\hatbbetacoop_n = \argmin_{\bbeta\in\mathbb{R}^p}
\frac{1}{2n}  \|\mathbf{y}- \mathbf{X}\bbeta \|^2
+ \lambda_n
\| \bbeta\|_{\mathrm{coop}}
,
\label{eqcooplassonormalized}
\end{equation}
where $\lambda_n = \lambda/n$.

\begin{theorem} \label{thmsupportconsistency}
If assumptions \textup{(A1)--(A5)} are satisfied, the coop-Lasso estimator is asymptotically
unbiased and has the property of exact support recovery:
%
\begin{equation}
\hatbbetacoop_n \inprob\bbeta^\star\quad\mbox{and} \quad
\prob \bigl(\supp(\hatbbetacoop_n) = \supp \bigr) \rightarrow1
, \label{eqconsistency}
\end{equation}
for every sequence $\lambda_n$ such that
$\lambda_n = \lambda_0
n^{- \gamma}, \gamma\in(0,1/2) $.
\end{theorem}

Compared to the group-Lasso, the consistency of support recovery for the
coop-Lasso differs primarily regarding possible intersection (besides inclusion
and exclusion) between groups and support.
This additional flexibility applies to every sign-coherent group.
Even if the support is the union of groups, when all groups are sign-coherent,
the coop-Lasso has still an edge on group-Lasso since the irrepresentable
condition \eqref{thICnoPnoN} is weaker.
Indeed, the norm in \eqref{thICnoPnoN} is dominated by the $\ell_2$
norm used
for the group-Lasso.
The next paragraph illustrates that this difference can have remarkable
outcomes.
Finally, when the support is the union of groups comprising
sign-incoherent ones,
there is no systematic advantage in favor of one or the other method.
While the
norm used by the coop-Lasso is dominated by the norm used by the group-Lasso,
the weighting matrix $\mathbf{D}$ has smaller entries for the latter.


\textit{Illustration.}
We generate data from the regression model \eqref{eqlinearreggroup}, with
$\bbeta^\star= (1,1,-1,-1,0,0,0,0)$, equipped with the group
structure  $\{\group\}_{k=1}^4 = \{\{1,2\},\allowbreak \{3,4\}, \{5,6\},\{
7,8\}\}$.
The vector $X$ is generated as a centered Gaussian random vector whose
covariance matrix $\bPsi$ is chosen so that the irrepresentable
conditions hold
for the coop-Lasso, but not for the group-Lasso, which, we recall, are more
demanding for the current situation, with sign-coherent groups.
The random error $\varepsilon$ follows a centered Gaussian
distribution with
standard deviation $\sigma=0.1$, inducing a very high signal to noise
ratio ($R^2=0.99$ on average), so that asymptotics provide a realistic
view of
the finite sample situation.

\begin{figure}

\includegraphics{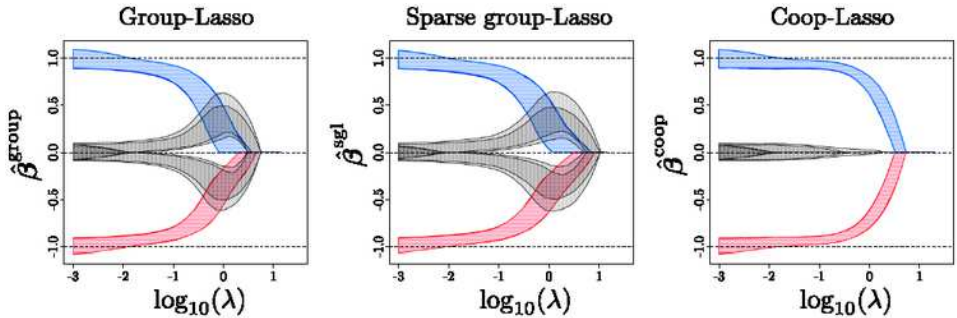}

\caption{50\% coverage intervals for the group (left), sparse group
(center) and (right) Lasso estimated coefficients along regularization
paths: coefficients from the support of $\bbeta^\star$ are marked by
colored horizontal stripes and the other ones by gray vertical stripes.}
\label{figconsistencyillustration}
\end{figure}

We generated 1000 samples of size $n=20$ from the described model, and
computed the
corresponding 1000 regularization paths for the group-Lasso, sparse group-Lasso
and coop-Lasso.
Figure~\ref{figconsistencyillustration} reports the 50\% coverage intervals
(lower and upper quartiles) along the regularization paths.
In this setup, the sparse group-Lasso behaves as the group-Lasso,
leading to
nearly identical graphs.
%
%
Estimation is difficult in this small sample problem
 ($n=20, p=8$), and the two
versions of the group-Lasso, which first select the wrong covariates, never
reach the situation where they would have a decisive advantage upon
OLS, while
the coop-Lasso immediately selects
the right covariates, whose coefficients steadily dominate the
irrelevant ones.
Model selection is also difficult, and the BIC criteria provided in
Section~\ref{secmodelselection} select often the OLS model (in about
$10\%$
and $50\%$ of cases for the coop-Lasso and the group-Lasso, respectively).
The average root mean square error on
parameters is of order $10^{-1}$ for all methods, with a slight edge
for the coop-Lasso.
The sign error is much more contrasted: $31\%$ for the coop-Lasso
\textit{vs.}
$46\%$ for the group-Lasso, not far better than the $50\%$ of OLS.


\section{Model selection}\label{secmodelselection}

Model selection amounts here to choosing the penalization parameter
$\lambda$,
which restricts the size of the
estimate $\hatbbeta(\lambda)$. Trial values
$\{\lambda_\mathrm{min},\ldots,\lambda_\mathrm{max}\}$ define the
set of models
we have to choose from along the regularization path.
The process aims at picking the model with minimum prediction error, or the
one closest to the model from which data have been generated,
assuming the model is correct, that is, equation~\eqref{eqlinearreggroup}
holds.
Here ``closest'' is typically measured by a distance between $\hatbbeta
$ and
$\bbeta^\star$, either based on the value of the coefficients or on their
support (true model selection), and sometimes also
on the sign correctness of each nonzero entry.

Among the prerequisite for the selection process to be valid,
the previous consistency analysis comes
up with suitable orders of magnitude for the penalty parameter $\lambda
$%
.
However, it does not provide a proper value to be plugged in
\eqref{eqcooplassolinear}
and the practice is to
use data driven approaches for selecting an appropriate penalty parameter.

Cross-validation is a recommended option [\citet{Hesterberg08}]
when looking for
the model minimizing the prediction error, but it is slow and not well
suited to
select the model closest to the true one.
Analytical criteria provide a faster way to perform model selection and,
though the information criteria AIC and BIC rely on asymptotic
derivations, they
often offer good practical performances.
The BIC and AIC criteria for the Lasso [\citet{2007ASZou}] and group-Lasso
[\citet{2006JRSSYuan}] have been defined through the effective
degrees of
freedom:
%
\begin{eqnarray}
\label{eqAICcoop}
\mathrm{AIC}(\lambda) & =& \frac{\|\mathbf{y} -
\hat{\mathbf{y}}(\lambda)\|^2}{\sigma^2}+ 2 \operatorname{df}(\lambda)
, \\
\label{eqBICcoop}
\mathrm{BIC}(\lambda) & =& \frac{\|\mathbf{y} -
\hat{\mathbf{y}}(\lambda)\|^2}{\sigma^2}+\log(n)
\operatorname{df}(\lambda)
,
\end{eqnarray}
where $\hat{\mathbf{y}}(\lambda)=\mathbf{X}\hatbbeta(\lambda)$ is
the vector of
predicted values for \eqref{eqcooplassolinear} with penalty parameter
$\lambda$,
$\sigma^2$ is the variance of the zero-mean Gaussian error variable
$\varepsilon$
in \eqref{eqlinearreggroup}
and $\operatorname{df}(\lambda)$ is the number of degrees of
freedom of the selected model. Assuming that
equation~\eqref{eqlinearreggroup} holds and a differentiability
condition on
the mapping $\hat{\mathbf{y}}(\lambda)$, \citet
{2004JASAEfron}, using Stein's
theory of unbiased risk estimate [\citet{1981ASStein}], shows that
%
\begin{equation}
\label{eqdefdf}
\operatorname{df}(\lambda) \doteq
\frac{1}{\sigma^2}\sum_{i=1}^n \operatorname{cov}(\hat{y}_i(\lambda),
y_i) =
\mathbb{E} \biggl[ \operatorname{tr} \biggl (\frac{\partial
\hat{\mathbf{y}}(\lambda)}{\partial\mathbf{y}} \biggr) \biggr]
,
\end{equation}
where the expectation is taken with respect to $\mathbf{y}$ or,
equivalently, to
the noise~$\boldsymbol\varepsilon$.
\citet{2006JRSSYuan} proposed an approximation of the trace term
in the
right-hand side of \eqref{eqdefdf}, which is used to estimate
$\operatorname{df}(\lambda)$ for the group-Lasso:
%
%
\begin{equation}
\label{eqdfgrouplasso}
\widetilde{\operatorname{df}}_\mathrm{group}(\lambda) = \sum_{k=1}^K
\1 \bigl( \|\hatbbetagroup_{\group}(\lambda) \|>0 \bigr)
 \biggl(1 +
\frac{ \|\hatbbetagroup_{\group}(\lambda) \|}
{ \|\bbetaols_{\group} \|}(p_k-1) \biggr)
,
\end{equation}
where $\1(\cdot)$ is the indicator function and $p_k$ is the number of elements
in $\group[k]$.
For orthonormal design matrices, \eqref{eqdfgrouplasso} is an unbiased
estimate of the true degrees of freedom of the group-Lasso and
\citet{2006JRSSYuan} suggest that this approximation is relevant
in more
general settings, by reporting that
``the performance of this approximate $C_p$-criterion [directly derived from~\eqref{eqdfgrouplasso}] is generally comparable with that of fivefold
cross-validation and is sometimes better.''

This approximation of $\operatorname{df}(\lambda)$ relies on the OLS
estimate and is
hence limited to setups where the latter exists and is unique. In particular,
the sample size should be larger than the number of predictors ($n \geq
p$). To
overcome this restriction,\vadjust{\goodbreak}
we suggest a more general approximation to the degrees of freedom,
based on the
ridge estimator
%
\begin{equation}
\label{eqbetaridge}
\hatbbetaridge(\gamma) =  (\mathbf{X}^\intercal \mathbf{X} +
\gamma\mathbf{I} )^{-1} \mathbf{X}^\intercal\mathbf{y}
,
\end{equation}
which can be computed even for small sample sizes ($n<p$).

\begin{proposition} \label{propdfcooplasso} Consider the
coop-Lasso estimator $\hatbbetacoop(\lambda)$ defined by~\eqref{eqcooplassolinear}. Assuming that data are generated according to
model \eqref{eqlinearreggroup}, and that $\mathbf{X}$ is orthonormal, the
following expression of $\widetilde{\operatorname{df}}_{\mathrm
{coop}}(\lambda)$ is
an unbiased estimate of $\operatorname{df}(\lambda)$ defined in \eqref
{eqdefdf} for
the coop-Lasso fit:
%
\begin{eqnarray}
\label{eqdfcooplasso}
\widetilde{\operatorname{df}}_{\mathrm{coop}}(\lambda) &=& \sum_{k=1}^K
\1 \bigl( \| (\hatbbetacoop_{\group}(\lambda) )^+ \|
>0 \bigr)
 \biggl(1 + \frac{p^k_+ -1}{1+\gamma}
\frac{  \| (\hatbbetacoop_{\group}(\lambda) )^+ \|}
{ \| (\hatbbetaridge_{\group}(\gamma) )^+ \|
} \biggr)
\nonumber
\\[-8pt]
\\[-8pt]
&&\hphantom{\sum_{k=1}^K}{} +
\1 \bigl( \| (\hatbbetacoop_{\group}(\lambda) )^- \|
>0 \bigr)
 \biggl(1 + \frac{p^k_- -1}{1+\gamma}
\frac{ \| (\hatbbetacoop_{\group}(\lambda) )^- \|}
{ \| (\hatbbetaridge_{\group}(\gamma) )^- \|} \biggr)
,
\nonumber\hspace*{-35pt}
\end{eqnarray}
%
where $p^k_+$ and $p^k_-$ are respectively the number of
positive and negative entries in $\hatbbetaridge_{\group}(\gamma)$.
\end{proposition}

Proposition~\ref{propdfcooplasso} raises a practical issue regarding
the choice of a good reference $\hatbbetaridge(\gamma)$. In
our numerous simulations (most of which are not reported here), we did not
observe a high sensitivity to $\gamma$, though high values degrade
performances.
When $\mathbf{X}$ is full rank we use $\gamma=0$ (the OLS estimate) and,
correspondingly, a~vanishing $\gamma$ (the Moore--Penrose solution) when
$\mathbf{X}$ is of smaller rank.
More refined strategies are left for future works.

Section~\ref{secnumerical} illustrates that, even in nonorthonormal settings,
plugging expression \eqref{eqdfcooplasso} for the degrees of freedom
$\operatorname{df}(\lambda)$ of the coop-Lasso in BIC~\eqref{eqBICcoop} or AIC
\eqref{eqAICcoop} provides sensible model selection criteria.
As expected, BIC, which is more stringent than AIC, is better at
retrieving the
sparsity pattern of $\bbeta^\star$, while AIC is slightly better regarding
prediction error.


\section{Simulation study}\label{secnumerical}

We report here experimental results in the regression setup, with the linear
regression model \eqref{eqlinearreggroup}.
Our simulation protocol is inspired from the one proposed by
\citeauthor{1995TBreiman} (\citeyear{1995TBreiman,1996ASBreiman}) to test the
nonnegative garrote estimator,
which inspired the Lasso.

\subsection{Data generation}\label{secnumericalgenerator}

The structure of $\bbeta^\star\in\mathbb{R}^p$ is controlled
through sparsity at
coefficient and group levels.
Here we have  $p=90$, forming $K=10$ groups of identical size, $p_k=9$.
All groups of parameters follow the same wave pattern: for $j\in\{
1,\ldots,9\}$,
$(\bbeta^\star_{\group})_j \propto\nu_k
 ( (h- |5-j | )^+ )^2$,
where $\nu_k\in\{0,1\}$ is a switch at the group level and $h\in\{
3,4,5\}$
governs the wave width, that is, the within-group sparsity, with respectively
$ |\supp_k |\in\{5,7,9\}$ nonzero coefficients in each group
included in the support
.
The covariates are drawn from a multivariate\vadjust{\goodbreak} normal distribution
$X \sim\mathcal{N}(\mathbf{0},\bPsi)$ with, for all $(j,j') \in
\{1,\ldots,p\}^2$, covariances $\Psi_{jj'} =
\rho^{|j-j'|}$, where $\rho\in[-1,1]$.
Finally, the response is corrupted by an error variable
$\varepsilon\sim\mathcal{N}(0,1)$ and the magnitude of the vector of
parameters
$\bbeta^\star$ is chosen to have an $R^2$ around $0.75$.

Note that the covariance of the covariates is purposely disconnected
from the
group structure. This setting may either be considered as unfair to the group
methods, or equally adverse for all Lasso-type estimators, in the sense that
none of their support recovery conditions are fulfilled when $\rho\neq0$.
Situations more or less advantageous for group methods are then
produced thanks to the parameter $h$, which determines how the support
of $\bbeta^\star$
matches the group structure.

\subsection{Results}

Model selection is performed with BIC~\eqref{eqBICcoop} for Lasso, group-Lasso
and coop-Lasso.
The estimation of the degrees of freedom for the Lasso is the number of nonzero
entries in $\hatbbetalasso(\lambda)$ [\citet{2007ASZou}].
As there is no such analytical estimate of the degrees of freedom for
the sparse
group-Lasso, we tested two alternative model selection strategies: standard
five-fold cross-validation (CV),
selecting the model with minimum cross-validation error, and the so-called
``1-SE rule'' [\citet{Breiman84}], which selects the most
constrained model whose
cross-validation error is within one standard error of the minimum.

\begin{figure}

\includegraphics{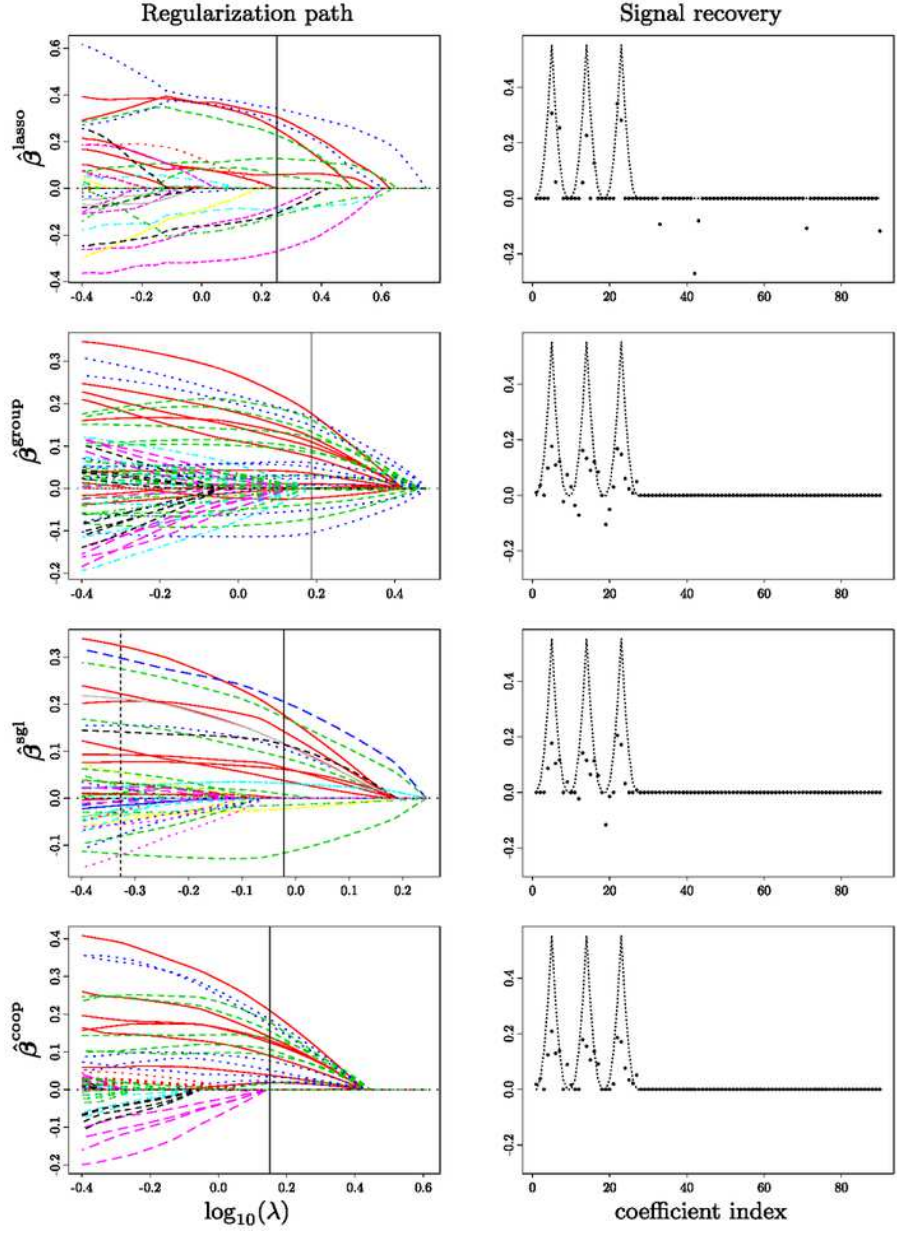}

\caption{Lasso, group, sparse group and coop Lasso estimates for a training
set of size $n=45$ drawn from the generation process of Section
\protect\ref{secnumericalgenerator}, with 3 active waves out of 10,
$ |\supp_k |/p_k=7/9$ and $\rho=0.4$.
Left: regularization paths, where each line type/color represents a
group of
parameters and the plain vertical line marks the model selected by the
1-SE rule for sparse group-Lasso and BIC otherwise;
right: true signal (dotted line) and estimated parameters for the selected
model (filled circles).}
\label{figbreimansignalexample}
\end{figure}

First, we display in Figure \ref{figbreimansignalexample} an example
of the
regularization paths obtained for each method for a small training set size
($n=p/2=45$) drawn from the model with three active groups having two zero
coefficients each ($ |\supp_k |=7$, $p_k=9$) and a moderate positive
correlation level ($\rho=0.4$).
%
%
As expected, the nonzero coefficients appear one at a time along the Lasso
regularization path and groupwise for the other methods, which detect the
relevant groups early, with some coefficients kept to zero for the sparse
group-Lasso and the coop-Lasso.
The sparse group-Lasso is qualitatively intermediate between the
group-Lasso and
the coop-Lasso, setting many parameters to zero, but keeping a few negative
coefficients in the solution.
The coefficients of the model estimated by BIC or the 1-SE rule are
displayed on
the right of each path. The Lasso estimate includes some nonzero coefficients
from irrelevant groups, but is otherwise quite conservative, excluding many
nonzero parameters from its support. This conservative trend is also observed
for the group methods, which exclude all irrelevant groups.
The three group estimates mostly agree on truly important coefficients, and
differ in the treatment of the spurious negative values that are
frequent for
group-Lasso, rarer for sparse group-Lasso and do not occur for coop-Lasso.

\begin{table}
\tabcolsep=0pt
\tablewidth=330pt
\caption{Average errors, with standard deviations, on
1000 simulations from the setup described in Section~\protect\ref{secnumericalgenerator}.
Each scenario differs in the number of observations $n$ and
the number of active variables per active group
$ |\supp_k |$ ($p_k=9$).
Sparse-cv and sparse-1-se designate the sparse group-Lasso with
$\lambda$
selected by cross-validation and by the 1-SE rule,
respectively}
\label{tabbreimanconvergence}
\vspace*{-3pt}
\begin{tabular*}{330pt}{@{\extracolsep{\fill}}lcccccc@{}}
\hline
& & \textbf{Lasso} & \textbf{Group} & \textbf{Sparse-cv} &
\textbf{Sparse-1-se} & \textbf{Coop}\\
\hline
Scenario && \multicolumn{5}{c@{}}{RMSE ($\times
10^3$)} \\
$ |\supp_k |=5$ & $n=45$ & 87.1 (0.5) & 95.0 (0.5) & 82.5 (0.5) & 88.1 (0.6) & 84.2 (0.5) \\
& $n=180$ & 43.7 (0.2) & 49.1 (0.2) & 41.7 (0.2) & 44.9 (0.2) & 43.5 (0.2) \\
& $n=450$ & 28.8 (0.1) & 33.4 (0.1) & 27.2 (0.1) & 30.9 (0.1) & 29.4 (0.1) \\
$ |\supp_k |=7$ & $n=45$  & 93.0 (0.5) & 85.8 (0.5) & 79.7 (0.4) & 83.6 (0.5) & 76.8 (0.5) \\
& $n=180$ & 48.4 (0.2) & 44.5 (0.2) & 42.2 (0.2) & 43.7 (0.2) & 40.4 (0.2) \\
& $n=450$ & 31.8 (0.1) & 30.3 (0.1) & 27.7 (0.1) & 30.0 (0.1) & 27.6 (0.1) \\
$ |\supp_k |=9$ & $n=45$ & 99.2 (0.4) & 82.0 (0.5) & 81.0 (0.4) & 83.2 (0.5) & 73.7 (0.5) \\
& $n=180$ & 52.5 (0.2) & 41.9 (0.2) & 43.3 (0.2) & 43.8 (0.2) & 39.0 (0.2) \\
& $n=450$ & 34.1 (0.1) & 28.7 (0.1) & 28.8 (0.1) & 30.6 (0.1) & 27.1 (0.1) \\
[4pt]
Scenario & & \multicolumn{5}{c@{}}{Mean sign error ($\%
$)}\\
$ |\supp_k |=5$ & $n=45$ & 13.8 (0.1) & 18.3 (0.2) & 36.7 (0.4) & 16.9 (0.3) & 13.3 (0.2) \\
& $n=180$ & \hphantom{0}8.4 (0.1) & 19.3 (0.2) & 36.1 (0.4) & 10.7 (0.2) & 13.0 (0.2) \\
& $n=450$ & \hphantom{0}6.1 (0.1) & 16.7 (0.2) & 35.5 (0.4) & \hphantom{0}7.1 (0.2) & 10.3 (0.2) \\
$ |\supp_k |=7$ & $n=45$ & 18.9 (0.1) & 12.9 (0.2) & 34.6 (0.4) & 16.8 (0.3) & 10.1 (0.2) \\
& $n=180$ & 11.9 (0.1) & 12.7 (0.2) & 34.5 (0.4) & 10.5 (0.2) & \hphantom{0}9.8 (0.2) \\
& $n=450$ & \hphantom{0}8.8 (0.1) & 10.4 (0.2) & 34.9 (0.4) & \hphantom{0}7.1 (0.2) & \hphantom{0}7.7 (0.2)
\\
$ |\supp_k |=9$ & $n=45$ & 24.4 (0.1) & \hphantom{0}8.1 (0.2) & 34.2 (0.4) & 17.3 (0.3) & \hphantom{0}7.9 (0.2) \\
& $n=180$ & 15.3 (0.1) & \hphantom{0}6.3 (0.2) & 33.5 (0.4) & 10.0 (0.2) & \hphantom{0}6.7 (0.2) \\
& $n=450$ & 11.2 (0.1) & \hphantom{0}4.3 (0.1) & 32.6 (0.4) & \hphantom{0}6.0 (0.2)& \hphantom{0}4.5 (0.1)
\\
\hline
\end{tabular*}\vspace*{-3pt}
\end{table}

Table \ref{tabbreimanconvergence} provides a more objective evaluation
of the
compared methods, based on the root mean square error (RMSE) and the support
recovery (more precisely, recovery of the sign of true parameters); prediction
error (not shown) is tightly correlated with RMSE in our setup.
Regarding the relative merits of the different methods, we did not
observe a
crucial role of the number of active groups and the covariate\vadjust{\goodbreak}
correlation level
$\rho$. We report results for a true support comprising 3 groups out
of 10 and
$\rho=0.4$, with various within-group sparsity and sample size scenarios.

All estimators perform about
equally in RMSE, the sparse group-Lasso with CV having a slight
advantage over
the coop-Lasso when many zero coefficients belong to the active groups,
and the
coop-Lasso being marginally but significantly better elsewhere.

Regarding support recovery, model selection with CV leads to models
overestimating the support of parameters. The 1-SE rule, which slightly harms
RMSE, is greatly beneficial in this respect.
BIC also performs very well,
incurring a very small loss due to model selection compared to the
oracle solution picking the model with best support recovery.
The Lasso dominates all the groups methods when many zero coefficients
belong to
the active groups. Elsewhere, group methods (with appropriate model selection
criteria) perform systematically significantly better for the small
sample sizes.
%
%
The coop-Lasso ranks first or a close second among group methods in all
experimental conditions.
It thus appears as the method of choice regarding inference issues when groups
conform to the sign-coherence assumption.
%

\subsection{Robustness}

The robustness to violations of the sign-coherence assumption is assessed
by switching a proportion $P_\sigma$ of signs in the vector~$\bbeta
^\star$, otherwise
generated as before. The sign of the corresponding covariates are switched
accordingly, to ensure that only the coop-Lasso estimators are affected
in the
process.

Table \ref{tabbreimanrobustness} displays the coop-Lasso RMSE that degrades
gradually with the amount of perturbation, becoming eventually worse
than the
Lasso, except for full groups.
%
%
\begin{table}
\tabcolsep=0pt
\tablewidth=335pt
\caption{Average errors, with standard deviations, on
1000 simulations from the setup of Table \protect\ref{tabbreimanconvergence} with
$n=180$, perturbed by switching a proportion $P_\sigma$ of signs in
$\bbeta^\star$}
\label{tabbreimanrobustness}
\begin{tabular*}{335pt}{@{\extracolsep{\fill}}lcccccc@{}}
\hline
&\multicolumn{3}{c}{\textbf{RMSE
($\boldsymbol{\times10^3}$)}} & \multicolumn{3}{c@{}}{\textbf{Mean sign error ($\boldsymbol\%$)}} \\[-5pt]
&\multicolumn{3}{c}{\hrulefill}&\multicolumn{3}{c@{}}{\hrulefill}\\
$\boldsymbol{P_{\sigma}}$ &  $ \boldsymbol{|\supp
_k |=5}$  &  $ \boldsymbol{|\supp
_k |=7}$  &
{$\boldsymbol{ |\supp_k |=9}$} &
{$ \boldsymbol{|\supp_k |=5}$} &
{$\boldsymbol{ |\supp_k |=7}$} &
{$\boldsymbol{ |\supp_k |=9}$} \\
\hline
0.1 & 46.9 (0.2) & 45.3 (0.2) & 45.8 (0.2) & 15.3 (0.2) & 12.4 (0.2) &
\hphantom{0}8.8 (0.2) \\
0.2 & 49.5 (0.3) & 48.9 (0.2) & 48.7 (0.2) & 17.8 (0.2) & 14.3 (0.2) &
\hphantom{0}9.8 (0.2) \\
0.3 & 51.0 (0.3) & 50.4 (0.3) & 50.4 (0.2) & 19.3 (0.2) & 14.8 (0.2) &
10.3 (0.2) \\
 0.4 & 51.6 (0.2) & 51.0 (0.2) & 50.2 (0.2) & 19.7 (0.2) & 14.8 (0.2) &
\hphantom{0}9.8 (0.2) \\
0.5 & 52.3 (0.3) & 51.3 (0.2) & 50.8 (0.2) & 20.0 (0.2) & 14.6 (0.2) &
\hphantom{0}9.3 (0.2) \\
\hline
\end{tabular*}
\end{table}
Regarding sign error, for small proportions of sign flip, the
coop-Lasso stays at
par with either Lasso or group-Lasso (see Table \ref{tabbreimanconvergence}),
but it eventually becomes significantly worse than both of them in most
situations. Thus, if the sign-coherence assumption is not firmly grounded,
either group-Lasso or its sparse version seem to be better options:
coop-Lasso only remains a second-best choice when there are less than
10\%
of sign mismatches within groups.


\section{Illustrations on real data}\label{secdata}

This section illustrates the applicability of the coop-Lasso on two
types of
predictors, that is, categorical and continuous covariates.
The first proposal may be widely applied to ordered categorical variables;
the second one is specific to microarray data, but should apply more generally
when groups of variables are produced by clustering.

In the first application, each group is formed by a set of variables
coding an ordered categorical variable.
Ordinal data are often processed
either by omitting the order property, treating them as nominal, or by
replacing each level with a
prescribed value, treating them as quantitative.
The latter procedure, combined with generalized linear regression,
leads to
monotone mapping from levels to responses.
Section \ref{secdataordinal} describes how coop-Lasso can bias the estimate
toward monotone mappings using a categorical treatment of ordinal variables.

In the second application of Section \ref{secdatacontinuous}, the
groups are
formed by continuous variables that are redundant noisy measurements (probe
signals) pertaining to a\vadjust{\goodbreak} common higher-level unobserved variable (gene
activity).
Sign-coherence is expected here, since each measurement should be positively
correlated with the activity of the common unobserved variable.
A similar behavior should also be anticipated when groups of
variables are formed by a clustering preprocessing step based on the Euclidean
distance, such as $k$-means or average linkage hierarchical clustering
[\citet{1998PNASEisen}; \citet{2006BSPark}; \citet
{2007BMCMa}].

\subsection{Monotonicity of responses to ordinal covariates}\label
{secdataordinal}

Monotonicity is easily dealt with by transforming ordinal covariates into
quantitative variables, but this approach is arbitrary and subject to many
criticisms when there is no well-defined numerical difference between levels,
which often lacks even for interval data when the lower or the upper
interval is
not bounded [\citet{Gertheiss09}].
Hence, the categorical treatment is often preferred, even if it fails
to fully
grasp the order relation.

The Lasso, group-Lasso or fused-Lasso have been applied to
the categorical treatment of ordinal features, with the aim to select
variables or aggregate adjacent levels [see \citet{Gertheiss10}
and references
within].
The coop-Lasso is used here to make a stronger usage of the order
relationship, by biasing
the mapping from levels to the response variable toward monotonic solutions.
Note that our proposal does not impose monotonicity and neither does it
prescribe an order (although several variations would be possible here).
In these respects, we depart from the approaches
imposing hard constraints on regression coefficients [\citet
{Rufibach20101442}].

\subsubsection{Methodology}

When not treated as numerical, ordinal variables are often coded by a
set of
variables that code differences between levels. Several types of
codings have
been developed in the ANOVA setting, with relatively little impact in the
regression setting, where the so-called dummy codings are intensively used.
Indeed, least squares fits
are not sensible to coding choices provided there is a one-to-one
mapping from
one to the other, so that codings only matter regarding the direct
interpretation
of regression coefficients.
However, codings evidently affect the solution in penalized
regression, and we will use here specific codings to penalize targeted
variations.
In order to build a monotonicity-based penalty, we simply use contrasts
that compare
two adjacent levels. An example of these contrasts is displayed in
Table~\ref{tabapplicationcontrasts}, with the corresponding codings,
known as
backward difference codings, which are simply obtained by solving a linear
system [\citet{Serlin85}].
%
%
\begin{table}
\tabcolsep=0pt
\tablewidth=205pt
\caption{Contrasts and codings for comparing
the adjacent levels of a covariate with 4 levels}\label{tabapplicationcontrasts}
\begin{tabular*}{205pt}{@{\extracolsep{\fill}}ld{2.0}d{2.0}d{2.0}d{4.0}d{4.0}d{4.0}@{}}
\hline
\textbf{Level} & \multicolumn{3}{c}{\textbf{Contrasts}} &
\multicolumn{3}{@{}c}{\textbf{Codings}} \\ \hline
0 & -1 & 0 & 0 & -3/4 & -1/2 & -1/4 \\
1 & 1 & -1 & 0 & 1/4 & -1/2 & -1/4 \\
2 & 0 & 1 & -1 & 1/4 & 1/2 & -1/4 \\
3 & 0 & 0 & 1 & 1/4 & 1/2 & 3/4 \\
\hline
\end{tabular*}
\end{table}
Note that several codings are possible for the contrasts given in
Table~\ref{tabapplicationcontrasts}. They differ in the definition of a
global reference level, whose effect is relegated to the intercept. As
we do not
penalize the intercept here, the particular choice has no outcome on the
solution.

Irrespective of the coding, group penalties act as a selection tool for factors,
that is, at variable level [\citet{2006JRSSYuan}].
On top of this, the sparse group penalty\vadjust{\goodbreak} usually presents the ability to
discard a level. With difference codings, some increments between adjacent
levels may be set to zero, that is, levels may be fused [\citet
{Gertheiss10}].
With the coop-Lasso penalty, all increments are urged to be sign-coherent,
thereby favoring monotonicity. As a side effect, level fusion may also be
obtained.

\subsubsection{Experimental setup}

We illustrate the approach on the Statlog ``German Credit'' data set [available
at the UCI machine learning repository, \citet{Frank10}], which gathers
information about people classified as low or high credit risks.
This binary response requires an appropriate model, such as logistic regression.
The coop-Lasso fitting algorithm is easily adaptable to generalized linear
models, following exactly the structure provided in Algorithm
\ref{algoactiveconstraint}, where the appropriate likelihood function replaces
the sum of square residuals in Step \ref{itemalgostep1}.

All quantitative variables
are used for the analysis, but we focus here on the regression
coefficients of
four variables, encoded as integers or nominal in the Statlog project, which
seem better interpreted as ordered nominal, namely:
\texttt{history}, with 4 levels describing the ability to pay back
credits in the past and now;
\texttt{savings}, with 4 levels giving the balance of the saving
account in currency intervals;
\texttt{employment}, with 5 levels reporting the duration of the
present employment in year intervals; and
\texttt{job}, with 4 levels representing an employment qualification scale.
Two other variables, related to the checking account status and
property, were
also encoded as nominal, but are not described here in full details
since they do not show distinct qualitative behaviors between methods.
We excluded from the ordinal variables categories merging two subcategories
possibly corresponding to different ranks,
such as
``critical account/other credits existing (not at this bank)'' in
\texttt{history}, or ``unknown/no savings account'' in \texttt{savings}.
For simplicity, we suppressed the corresponding examples, thus ending
with a~total of 330 observations, split into three equal-size learning,
validation and test sets. We estimate the logistic regression
coefficients on
the learning set, perform model selection from deviance or
misclassification error on
the validation set, and finally keep the test set to estimate
prediction performances.\looseness=1

\subsubsection{Results}

\begin{figure}

\includegraphics{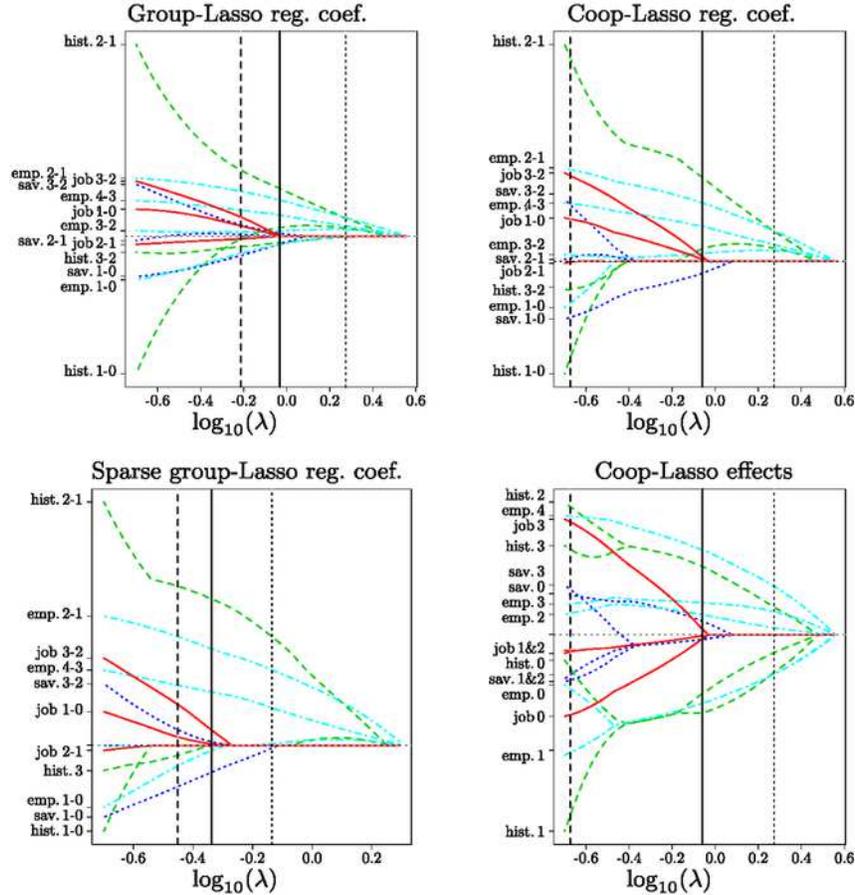}

\caption{Regularization paths for four ordinal covariates (history, savings,
job and employment) for the group, coop, and sparse group-Lasso on
the contrast coefficients obtained from backward difference coding
(top left, top right and bottom left, respectively). The
transcription of contrasts to levels is also displayed for coop-Lasso
(bottom right). The vertical lines mark the model selected by
cross-validation on the validation set, for different criteria:
deviance (plain), misclassification rate (dashed), and weighted
misclassification error (dotted).}
\label{figgermancreditdatapath}
\end{figure}

The performances of the three group methods are identical, either
evaluated in terms of\vadjust{\goodbreak}
deviance, classification error rate or weighted misclassification (unbalanced
misclassification losses are provided with the data set).
The regression coefficients differ, however, as shown in Figure
\ref{figgermancreditdatapath} displaying the regularization paths for all
methods.
%
%
Recall that we only represent the ordinal covariates \texttt{history},
\texttt{savings}, \texttt{employement} and \texttt{job}. Each
coefficient represents
the increment between two adjacent levels, with positive and negative values
resulting in an increase and decrease, respectively.
Monotonicity with respect to all levels is reached if all the values
corresponding to a factor are nonnegative or nonpositive.
We also provide an alternative view of the coop-Lasso path, with the overall
effects corresponding to levels, obtained by summing up the increments.

Most factors are not obviously amenable to quantitative coding since
there is no natural distance between levels, but we, however, underline that
using the usual quantitative transformation with equidistant values
followed by
linear regression would correspond here to identical increments between levels.
Obviously, all displayed solutions radically contradict this linear
trend hypothesis.

Our three solutions differ regarding monotonicity, which is almost never
observed along the group-Lasso regularization path. The sparse
group-Lasso paths
have long sign-coherent sections, where group-Lasso infers slight
wiggles. These
sections extend further with the coop-Lasso.
However, as the coop penalty goes to zero, sign-coherence is no longer
preserved, and all methods eventually reach the same solution.

The sparse group and the coop-Lasso set some increments to zero,
leading to the
fusion of adjacent levels that should be welcomed regarding interpretation.
The solutions tend to agree on these fusions on long sections of the
paths, with
some additional fusions of the sparse group-Lasso when slight monotonic
solutions are provided by the coop-Lasso (see \texttt{employment},
levels 2 and
3, and \texttt{savings} levels 1 and 2).
These fusions are perceived more directly on the coop-Lasso path of effects,
displayed in the bottom right of Figure~\ref{figgermancreditdatapath}, where
the effect of each level is displayed directly.

\subsection{Robust microarray gene selection}\label{secdatacontinuous}

Most studies on response to che\-motherapy have considered breast cancer
as a single homogeneous entity. However, it is a complex disease
whose strong heterogeneity should not be overlooked. The data set
proposed by \citet{2006JCOhess} consists in gene expression
profiling of patients treated with chemotherapy prior to surgery,
classified as presenting either a pathologic complete response (pCR) or a~residual disease (not-pCR). It records the signal of 22,269
probes\footnote{%
Actually, the data set reports the average signal in probe sets, which
are a
collection of probes designed to interrogate a given sequence. In this paper
the term ``probe'' designates Affymetrix probe sets to avoid confusion
with the
group structure that will be considered at a higher level.}
examining the human genome, each probe being related to a unique gene.
Following \citet{2011JSFDSjeanmougin}, we restrict our analysis
to the basal
tumors:
for this particular subtype of breast cancer, clinical and pathologic features
are homogeneous in the data set, whereas the response to chemotherapy is
balanced, with 15 tumors being labeled pCR and 14 not-pCR.
This setup is thus propitious to the statistical analysis of response to
chemotherapy from the sole activity of genes.

\subsubsection{Methodology}

The usual processing of microarray data relies on probe
measurements\vadjust{\goodbreak}
that are related to genes in the final interpretation of
the statistical analysis. Here we would like to take a different
stance, by
gathering all the measurements associated to gene entities at an early
stage of
the statistical inference process.
As a matter of fact, we typically observe that some probes related to
the very
same gene have different behaviors.
Requiring a consensus at the gene level supports biological coherence, thus
exercising caution in an inference process where statistically
plausible explanations are numerous, due to the noisy probe signals and
to the
cumbersome $n \ll p$ setup (here $n=29$ and $p=22\mbox{,}269$).
Since the probes related to a given gene relate to sequences that are
predominantly cooperating, the sign-coherence assumed by the coop-Lasso is
particularly appropriate to improve robustness to the measurement noise
and to
encourage biologically plausible solutions.

Our protocol includes a preselection of probes that facilitates the
analysis for
the nonadaptive penalization methods compared here, and also provides an
assessment of the benefits of adding seemingly less relevant
probes into the statistical analysis. We proceed as follows:
\begin{itemize}
\item select a restricted number $d$ of probes from classical differential
analysis, where probes are sorted by increasing $p$-values;
\item determine the genes associated to these $d$ probes, retrieve all the
probes related to these genes, and select the corresponding $p$ probes,
$p \ge d$,
regardless of their signal;
\item fit a model with group penalties where groups are defined by genes.
\end{itemize}

\subsubsection{Experimental setup}

We select the first $d=200$ most differentiated probes, as identified
by the analysis of \citet{2011JSFDSjeanmougin}, on the 22,269 probes
for the $n=29$ patients with basal tumor. These 200 probes correspond
to 172 genes, themselves associated to $p=381$ probes on the
microarray as a whole, with 1 to 13 probes per gene. We clearly enter
the high-dimensional setup with $p > 13 \times n$.

All signals are
normalized to have a unitary within-class variance. We compare
then the Lasso on the $d=200$ most differentiated probes, with the
Lasso and group, sparse group and coop Lasso on the $p=381$ probes.
All fits are produced with our code (available at
\texttt{\href{http://stat.genopole.cnrs.fr/logiciels/scoop}{http://stat.genopole.cnrs.fr/}
\href{http://stat.genopole.cnrs.fr/logiciels/scoop}{logiciels/scoop}}).

Well-motivated analytical model selection criteria are not available
today for Lasso-type penalties beyond the regression setup. Here,
model selection is carried out by 5-fold cross-validation: we
evaluate the $\mathrm{CV}$ error for each method with the same block
partition using either the binomial deviance or the unweighted
classification error. 

\subsubsection{Results}

The 5-folds $\mathrm{CV}$ scores, either based on deviance or
misclassification losses,\vadjust{\goodbreak} are reported for each estimation method in
Table \ref{tabcvresults}, which also displays the number of selected groups
and features for the models selected by minimizing the CV score.
%
\begin{table}
\tabcolsep=0pt
\caption{$\mathrm{CV}$ scores for misclassification error and
binomial deviance on the basal tumor data.
The minimizer of $\mathrm{CV}$ for misclassification and deviance
are respectively denoted by $\lambdaclass$ and~$\lambda^{\mathrm{dev}}$; the number of selected groups and features
respectively refers to genes and probes}
\label{tabcvresults}
\begin{tabular*}{\textwidth}{@{\extracolsep{\fill}}lcccccc@{}}
\hline
& & \textbf{Probes} & \textbf{Lasso} & \textbf{Group} &
\textbf{Sparse} & \textbf{Coop} \\
\hline
Model selection rule &&
\multicolumn{5}{c@{}}{CV score $\times100$
(standard error)} \\
Classification
& $\lambdaclass$ & 10.3 (5.8)\hphantom{0} & 6.9 (4.9) & \hphantom{0}3.4 (3.5) & 3.4 (3.5)
& \hphantom{0}3.4 (3.5) \\
Deviance &
$\lambda^{\mathrm{dev}}$ & 76.5 (37.6) & 67.2 (32.3) & 13.7 (8.1) & 20.5 (10.0) & 13.8 (7.9) \\[4pt]
Model selection rule &&
\multicolumn{5}{c@{}}{\# selected groups (features)}\\
Classification
& $\lambdaclass$ & 17 (17) & 16 (17) & 11 (15) & 14 (21)
& \hphantom{0}9 (11) \\
Deviance
& $\lambda^{\mathrm{dev}}$ & 19 (19) & 17 (18) & 13 (21) & 16 (26)
& 14 (18) \\
\hline
\end{tabular*}
\end{table}

Expanding the set of probes from $d$ to $p$ slightly improves the performances
of the Lasso, and considerable further progresses are brought by all group
methods, which misclassify about 1 patient among the 29 and quarter deviance
scores.\footnote{%
A note of caution regarding performances: scores comparisons are
fair here, in the sense that the $\mathrm{CV}$ scores are optimized with
respect to a single parameter $\lambda$, whose role is analog for
all. Additional simulations (not reported here) show that, for all group
methods, the
${\mathrm{CV}}$ error is stable with respect to the random choice of
folds and that the $\mathrm{CV}$ curves are smooth around their
minima. However, the minimizers of
$\mathrm{CV}$ are biased estimates of out-of-sample scores, and the
representativeness of their observed difference
can be questioned.}
As expected, less genes are selected by group methods;
the difference is more important for the minimizers of the
misclassification score, and, among those, for the group-Lasso and coop-Lasso
that comply more stringently to the group structure.
These observations indicate that the group structure defined by genes provides
truly useful guidelines for inference.

\begin{figure}

\includegraphics{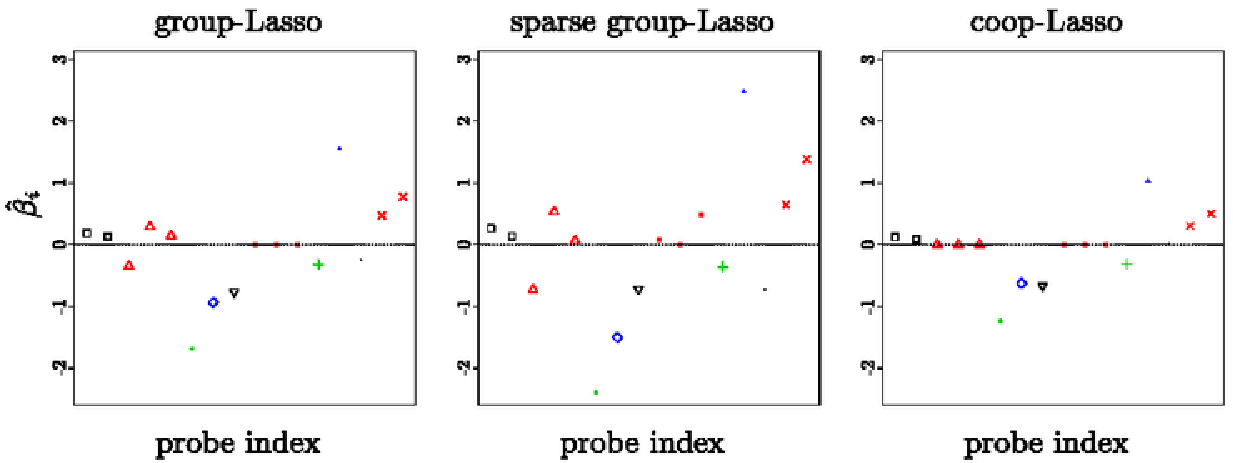}

\caption{Logistic regression coefficients attached to each probe for
group, sparse group and coop Lasso. Each marker (color and
symbol) designates the gene associated to the probe:
\textsc{rnps1}
\mbox{\emph{(}\protect\includegraphics{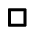}\emph{)}},
\textsc{msh6}
\mbox{\emph{(}\protect\includegraphics{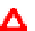}\emph{)}},
\textsc{prps2}
\mbox{\emph{(}\protect\includegraphics{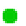}\emph{)}},
\textsc{h1fx}
\mbox{\emph{(}\protect\includegraphics{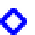}\emph{)}},
\textsc{mfge8}
\mbox{\emph{(}\protect\includegraphics{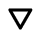}\emph{)}},
\textsc{sulf1}
\mbox{\emph{(}\protect\includegraphics{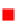}\emph{)}},
\textsc{rnf115}
\mbox{\emph{(}\protect\includegraphics{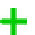}\emph{)}},
\textsc{rnf38}
\mbox{\emph{(}\protect\includegraphics{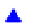}\emph{)}},
\textsc{thnsl2}
\mbox{\emph{(}\protect\includegraphics{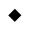}\emph{)}}
and \textsc{edem3}
\mbox{\emph{(}\protect\includegraphics{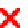}\emph{)}}.}
\label{figprobes}
\end{figure}

The sparsity numbers differ among the group methods, coop-Lasso
selecting as
many genes as group-Lasso and fewer probes, and sparse group-Lasso retaining
slightly more genes and probes.
A more detailed picture is provided in Figure~\ref{figprobes}, which shows
the regression coefficients for the three group estimators adjusted on
the whole
data set with their respective $\lambdaclass$ values.
%
Among the three methods, a~total of 15 groups (i.e., genes) are selected.
For readability, we only represent the 10 leading groups of regression
coefficients (according to their average norm).
We first oberve that the magnitude of coefficients differs for each
method, the
coop-Lasso having the smallest one. In fact, there is a wide range of
$\lambdaclass$ values for which the miclassification score is minimal\vadjust{\goodbreak}
for the
coop-Lasso, enabling to choose a highly penalized solution without affecting
accuracy.
The magnitude apart, the group methods have qualitatively the same
behaviors for
all unitary groups but one, with \textsc{thnsl2}
(\includegraphics{520i09.eps}) being set to zero
by the
coop-Lasso.
The same patterns are observed for two other groups,
\textsc{rnps1} ({\includegraphics{520i01.eps}}) and
\textsc{edem3} ({\includegraphics{520i10.eps}}), whose
regression coefficients are consistently estimated to be sign-coherent.
Then, \textsc{sulf1} ({\includegraphics{520i06.eps}}),
though being estimated sign-coherent by the sparse group-Lasso, is
excluded from
the support of the group and coop Lasso.
Finally, \textsc{msh6} ({\includegraphics{520i02.eps}}),
estimated as sign incoherent with the two groups methods, is excluded
from the
support for the coop-Lasso.

Overall, the probe enrichment scheme we propose here leads to considerable
improvements in prediction performance. This better statistical
explanation is
obtained without impairing interpretability, since sign-coherence is actually
often satisfied by all methods and strictly enforced by the coop-Lasso.
As often in this type of study, several methods provided similar
prediction performances, but the explanation provided by the coop-Lasso is
simpler, both from a statistical and from a biological viewpoint. Note
that the
coefficient paths (not shown) diverge early between the group and coop methods,
so that the above-mentioned discrepancies are not simply due to model selection
issues.
As a final remark, we observed qualitatively similar behaviors when
the initial number of probes $d$ ranged from 10 to 2000.
For $d \leq1000$, the group methods always performed best, with
approximately identical
classification errors, the group-Lasso and coop-Lasso slightly
dominating the
sparse group-Lasso in terms of deviance. With larger initial sets of probes,
the enrichment procedure becomes less efficient, and all methods provide
similar decaying results. The chosen setup displayed here, with
$d=200$, leads to the
smallest classification error for all methods, and was chosen for being
representative of the most interesting regime.\vadjust{\goodbreak}




\section{Discussion}\label{secdiscussion}

The \emph{coop-Lasso} is a variant of the \emph{group-Lasso} that was
originally proposed in the context of multi-task learning, for
inferring related networks with Gaussian Graphical Models
[\citet{2010SCChiquet}]. Here we develop its analysis in the linear
regression setup and demonstrate its value for prediction and inference with
generalized linear models.
Along with this paper we provide an implementation of the fitting
algorithm in the
\texttt{R} package \texttt{scoop}, which makes this new penalized estimate
publicly available for linear and logistic regression (the \emph
{coop-Lasso} for
multiple network inference is also available in the \texttt{R} package
\texttt{simone}).

The coop-Lasso differs from the group-Lasso and sparse group-Lasso
[Fried\-man, Hastie and
Tibshirani (\citeyear{2010preprintFriedman})] by the assumption that the group
structure is \emph{sign-coherent}, namely, that groups gather either
nonpositive, nonnegative or null parameters, enabling the recovery of various
within-group sign patterns (positive, negative, null, nonpositive,
nonnegative, nonnull).
This flexibility greatly reduces the incentive to drive within-group
sparsity with an additional parameter that later
leads to an unwieldy model selection step.
However, the relevance of the sign-coherence assumption should be firmly
established since it plays an essential role in the performance of
coop-Lasso compared to the sparse group-Lasso.

Under suitable irrepresentable conditions, the proposed penalty leads
to consistent model selection, even when the true sparsity pattern
does not match the group structure. When the groups are sign-coherent
the coop-Lasso compares favorably to the group-Lasso, recovering the
true support under the mildest assumptions.

We present an approximation of the effective degrees of freedom of the
coop-Lasso which, once plugged into AIC or BIC, provides a fast way to select
the \mbox{tuning} parameter in the linear regression setup.
We provide empirical results demonstrating the capabilities of the
coop-Lasso in
terms of prediction and parameter selection, with BIC performing very well
regarding support recovery even for small sample sizes.

We illustrate the merits of the coop-Lasso applied to the analysis to ordinal
and continuous predictors. With
an apposite coding, such as forward or backward
difference coding, the sign-coherence assumption is transcribed
in a monotonicity assumption, which does not require to stipulate the
usual and controversial mapping from levels to quantitative variables.
Finally, the application to genomic data opens a vast potential field
of great
practical interest for this type of penalty, both in terms of
prediction and
interpretability.
Our forthcoming investigations will aim at substantiating this ambition by
conducting large scale experiments in this application domain.


\begin{appendix}
\section*{Appendix: Proofs}\label{appm}

\subsection{\texorpdfstring{Proof of Lemma~\protect\ref{lemsubdifferential}}{Proof of Lemma 1}}

Let us use
${\cal T}_k$ as a shorthand for ${\cal S}_k(\bbeta)$, \citet
{2010SCChiquet} show that the
subdifferential $\btheta$ obey the following conditions:
%
\begin{subequation}
\label{eqsubgradient}
%
\begin{eqnarray}\label{eqsubgradienta}
 &\displaystyle \max (\norm{\btheta_{\group}^+},\norm{\btheta_{\group
}^-}  )\leq w_k
   \qquad \mbox{if } \bbeta_{\group} = \mathbf{0}  ,&
\\\label{eqsubgradientb}
 &\displaystyle \btheta_{{\cal T}_k} = \frac{w_k  \bbeta_{{\cal T}_k}}{\norm{\bbeta_{{\cal
 T}_k}}},\qquad
\norm{\btheta_{{\cal T}_k^c}^-} \leq w_k, \qquad
\norm{\btheta_{{\cal T}_k^c}^+} = 0&\nonumber
\\[-8pt]
\\[-8pt]
\eqntext{\mbox{if } \|\bbeta_{\group}^+\|>0,  \|\bbeta
_{\group}^-\|=0 ,}
\\\label{eqsubgradientc}
 &\displaystyle \btheta_{{\cal T}_k} = \frac{w_k  \bbeta_{{\cal T}_k}}{\norm{\bbeta_{{\cal
 T}_k}}},\qquad
\norm{\btheta_{{\cal T}_k^c}^+} \leq w_k, \qquad
\norm{\btheta_{{\cal T}_k^c}^-} = 0
&\nonumber
\\[-8pt]
\\[-8pt] \eqntext{\mbox{if } \|\bbeta_{\group}^-\|>0,  \|\bbeta
_{\group}^+\|=0 ,}
\\
\label{eqsubgradientd}
 &\displaystyle \forall j\in\group    \qquad   \btheta_{j} = w_k  \bbeta_{j} \norm{\sign(\beta_j)\bbeta}^{-1}
&\nonumber
\\[-8pt]
\\[-8pt]   \eqntext{\mbox{if } \|\bbeta_{\group}^-\|>0,  \|\bbeta
_{\group}^+\|>0  .}
\end{eqnarray}
\end{subequation}
We thus simply have to prove the equivalence of conditions
\eqref{eqsubgradientcompact} and \eqref{eqsubgradient} for all
$\bbeta_{\group}$
values.

For $\bbeta_{\group}=\mathbf{0}$, \eqref{eqsubgradientcompact} reads
%
\begin{equation}
\norm{\btheta_{\group}^+} \leq w_k  \quad \mbox{and} \quad
\norm{\btheta_{\group}^-} \leq w_k
,
\label{eqsubgradientabis}
\end{equation}
which is equivalent to \eqref{eqsubgradienta}.

For $\bbeta_{\group}\neq\mathbf{0}$, the equalities for $\btheta
_{{\cal T}_k}$ in
\eqref{eqsubgradientb}--\eqref{eqsubgradientd} are equivalent to
\eqref{eqsubgradientcompactb}, thus setting the equivalence between
\eqref{eqsubgradientcompact} and \eqref{eqsubgradient} for all nonzero
coefficients.
For $\bbeta_{{\cal T}_k^c}$, let us consider the case~\eqref{eqsubgradientb},
where all nonzero parameters within group $k$ are positive.
The first equation of~\eqref{eqsubgradientb} implies that $\norm{\btheta_{{\cal T}_k}^+} = w_k$
and $\norm{\btheta_{{\cal T}_k}^-} = 0$.
Hence, $\norm{\btheta_{{\cal T}_k^c}^-} \leq w_k$ and
$\norm{\btheta_{{\cal T}_k^c}^+}=0$ imply \eqref
{eqsubgradientabis}, so that
\eqref{eqsubgradientb} implies~\eqref{eqsubgradientcompact}.
The contraposition is also easy to check. From \eqref{eqsubgradientcompactb},
when all coefficients are positive, we have that $\norm{\btheta
_{{\cal T}_k}^-}=0$
and $\norm{\btheta_{{\cal T}_k}^+} = w_k$.
Then, this implies that \eqref{eqsubgradientcompacta} reads
\[
\norm{\btheta_{{\cal T}_k^c}^-} \leq w_k
 \quad \mbox{and} \quad
\norm{\btheta_{{\cal T}_k^c}^+}=0
,
\]
which defines $\btheta_{{\cal T}_k^c}$ in \eqref{eqsubgradientb}. The
proof is similar
for \eqref{eqsubgradientc} where all nonzero parameters within group
$k$ are
positive.

\subsection{\texorpdfstring{Proof of Proposition~\protect\ref{propdfcooplasso}}{Proof of Proposition 1}}

We assume here that $\mathbf{X}^\intercal \mathbf{X} = \mathbf{I}_p$.
We introduce the ridge estimator in the computation of the trace in
equation~\eqref{eqdefdf}, through the chain rule, yielding an unbiased
estimate of $\operatorname{df}$:
\begin{eqnarray*}
\widetilde{\operatorname{df}}_{\mathrm{coop}}(\lambda)
&= &\operatorname{tr} \biggl(\frac{\partial\hat{\mathbf{y}}(\lambda
)}{\partial\mathbf{y}} \biggr)
=   \operatorname{tr} \biggl(
\frac{\partial\mathbf{X}^{\intercal}\hatbbetacoop(\lambda)}
{\partial\hatbbetaridge(\gamma)}
\,\frac{\partial\hatbbetaridge(\gamma)}{\partial\mathbf{y}}
\biggr) \\
&= &  \frac{1}{1 + \gamma} \sum_{k=1}^K \sum_{j\in\group} \frac
{\partial
\hatbetacoop_j(\lambda)}{\partial\hatbetaridge_j(\gamma)}
,
\end{eqnarray*}
where the last equation derives from the definition \eqref
{eqbetaridge} of the
ridge estimator with regularization parameter $\gamma$. Then, the
expression of
the coop-Lasso as a function of the ridge regression estimate is simply obtained
from equation~\eqref{eqcooportho}, using that, in the orthonormal
case, we have $\hatbbetaols= (1+\gamma)
\hatbbetaridge(\gamma)$. Dropping the reference to $\lambda$ and
$\gamma$ that
is obvious from the context, we have, $\forall k \in\{1,\ldots,K\}$ and
$\forall j\in\group$,
%
%
\begin{equation}
\label{eqcoopridgeortho}
\hatbeta_j^{\mathrm{coop}} = \biggl (
1-\frac{\lambda
w_k}{(1+\gamma) \|\bvarphi_j(\hatbbetaridge_{\group}) \|}
 \biggr)^{  +}
(1+\gamma) \hatbeta_j^{\mathrm{ridge}}
.
\end{equation}
Then, for $j\in\group$, routine differentiation gives
\begin{eqnarray*}
\frac{1}{1+\gamma}\,\frac{\partial\hatbetacoop_j}{\partial
\hatbetaridge_j} &=  &
\1 (\|\hatbetacoop_j\|>0 ) \\
&&{}   \times \biggl(1 -
\frac{\lambda w_k} {(1+\gamma)}  \biggl(
\frac{1}{\|\bvarphi_j(\hatbbetaridge_{\group})\|} -
\frac{(\hatbetaridge_j)^2}{\|\bvarphi_j(\hatbbetaridge_{\group})\|^3}
 \biggr)
 \biggr)
.
\end{eqnarray*}
The summation over the positive and negative elements of $\group$
reduces to two
terms
\begin{eqnarray*}
&&\frac{1}{1+\gamma} \sum_{j\in\group}\frac{\partial\hatbetacoop
_j}{\partial
\hatbetaridge_j}\\
&& \qquad =   \1 \bigl(\|(\hatbbetacoop_{\group})^+\|>0 \bigr)
\biggl(p^k_+  -
\frac{\lambda w_k}{1+\gamma}
\frac{(p^k_+ -1)}{\|(\hatbbetaridge_{\group})^+\|}  \biggr)
\\ && \qquad  \quad {}+
\1 \bigl(\|(\hatbbetacoop_{\group})^-\|>0 \bigr)  \biggl(p^k_-  -
\frac{\lambda w_k}{1+\gamma}
\frac{(p^k_- -1)}{\|(\hatbbetaridge_{\group})^-\|}  \biggr) \\
&& \qquad =    \1 \bigl(\|(\hatbbetacoop_{\group})^+\|>0 \bigr) +
\biggl (1 - \frac{\lambda w_k}{(1+\gamma)\|(\hatbbetaridge_{\group
})^+\|}  \biggr)^+ (p^k_+ -1)
\\ && \qquad  \quad {}+ \1 \bigl(\|(\hatbbetacoop_{\group})^-\| >0 \bigr) + \biggl (1 -
\frac{\lambda w_k}{(1+\gamma)\|(\hatbbetaridge_{\group})^-\|}
 \biggr)^+(p^k_- -1)
.
\end{eqnarray*}
From \eqref{eqcoopridgeortho}, we have, $\forall k \in\{1,\ldots,K\}
$ and $\forall j\in\group ,
$
\[
 \biggl(
1-\frac{\lambda w_k}
{(1+\gamma) \|\bvarphi_j(\hatbbetaridge_{\group}) \|}
 \biggr)^{  +}
= \frac{1}{1+\gamma} \frac{ \|\bvarphi_j(\hatbbetacoop
_{\group}) \|}{ \|\bvarphi_j(\hatbbetaridge_{\group
}) \|}
,
\]
which is used twice to simplify the previous expression. Summing over all
groups concludes the proof.

\subsection{\texorpdfstring{Proof of Theorem \protect\ref{thmsupportconsistency}}{Proof of Theorem 2}}

Our asymptotic results are established on the scaled problem~\eqref
{eqcooplassonormalized}.\vadjust{\goodbreak}
We then follow the three steps proof technique proposed by \citet
{2007JRSSYuan} for
the Lasso and also applied by \citet{2008JMLRBach} for the group-Lasso:
\begin{longlist}[(3)]
\item[(1)] restrict the estimation problem to the true support;
\item[(2)] complete this estimate by 0 outside the true support;
\item[(3)] prove that this artificial estimate satisfies optimality
conditions for the original coop-Lasso problem with probability
tending to 1.
\end{longlist}
Then, under (A2), the solution is unique, leading to the conclusion
that the
coop-Lasso estimator is equal to this artificial estimate with
probability tending to 1, which ends the proof. Note, however, a slight
yet important difference along the discussion: since we authorize
divergences between the group structure $\{\group\}_{k=1}^K$ and the true
support $\supp$, the irrepresentable conditions (A4)--(A5) for the
coop-Lasso cannot
be expressed simply in terms of coop-norms [as it is done with the
group-norm in \citet{2008JMLRBach}].
We will see that this does not impede the development of the proof.

As a first step, we prove two simple lemmas. Lemma
\ref{lemsupportconsistency} states that the coop-Lasso estimate,
restricted on
the true support $\supp$, is consistent when $\lambda_n \to0$. Lemma
\ref{lemUseIrrepresentableCond} provides the basis for the inequalities
\eqref{thICnoPnoN} and \eqref{thICPorN} that express our irrepresentable
conditions.

\begin{lemma} \label{lemsupportconsistency} Assuming \textup{(A1)--(A3)}, let
$\tildebbeta_\supp^n$ be the unique minimizer of the regression
problem restricted to the true support $\supp$:
\[
\tildebbeta_\supp^n = \argmin_{\mathbf{v} \in\Rset^{|\supp|}}
\frac{1}{2} \|\mathbf{y} - \mathbf{X}_{\centerdot\supp}\mathbf
{v}\|_n^2 +
\lambda_n \sum_{k\dvtx\supp_k \neq\varnothing} w_k (\|\mathbf{v}_{\supp
_k}^+\| +
\|\mathbf{v}_{\supp_k}^-\|)
,
\]
where $\|\cdot\|_n = \|\cdot\|/n$ denotes the empirical norm.

If $\lambda_n \rightarrow0$, then $\tildebbeta_\supp^n \inprob
\bbeta^\star_{\supp}.$
\end{lemma}

\begin{pf}
This lemma stems from standard results of M-estimation
[\citet{1998CUPVaart}].
Let $\boldsymbol\varepsilon= \mathbf{y} - \mathbf{X}\bbeta^\star
$, and write
$\bPsi^n = \mathbf{X}^\intercal\mathbf{X}/n$.
If $\lambda_n \rightarrow0$, then under (A1)--(A2), for any $\mathbf{v}
\in
\Rset^{|\supp|}$
\begin{eqnarray*}
Z_n(\mathbf{v}) & =&
\frac{1}{2} \|\mathbf{y} -
\mathbf{X}_{\centerdot\supp}\mathbf{v}\|_n^2 + \lambda_n \sum
_{k\dvtx \supp_k \neq
\varnothing} w_k (\|\mathbf{v}_{\supp_k}^+\| + \|\mathbf{v}_{\supp
_k}^-\|) \\
&=& \frac{1}{2}
(\bbeta^\star_{\supp}-\mathbf{v})^\intercal\bPsi_{\supp\supp}^n
(\bbeta^\star_{\supp}-\mathbf{v}) -
\frac{1}{n} \boldsymbol\varepsilon^\intercal\mathbf{X}_{\centerdot
\supp}(\bbeta^\star_{\supp}-\mathbf{v}) +
\frac{\boldsymbol\varepsilon^\intercal \boldsymbol\varepsilon
}{2n} \\
&&{}   +  \lambda_n \sum_{k,
\supp_k \neq\varnothing} w_k (\|\mathbf{v}_{\supp_k}^+\| + \|\mathbf
{v}_{\supp_k}^-\|)
\end{eqnarray*}
tends in probability to
\[
Z(\mathbf{v}) =
\tfrac{1}{2}(\bbeta^\star_{\supp}-\mathbf{v})^\intercal\bPsi
_{\supp\supp} (\bbeta^\star_{\supp}-\mathbf{v}) +
\tfrac{1}{2}\sigma^2.
\]
It follows from the strict convexity of $Z_n$ that  $\argmin Z_n(\mathbf{v})
\inprob\argmin Z(\mathbf{v}) = \bbeta^\star_{\supp}$ [\citet
{2000ASKnight}], which ends the
proof.\vadjust{\goodbreak}~%
\end{pf}

\begin{lemma} \label{lemUseIrrepresentableCond}
Consider a sequence of random variables $S_n$ such that
 $S_n \inprob S$. Suppose there exists $\delta>0$ such that for
a given norm
$\mu$ the limit $S$ is bounded away from 1:
\[
\mu(S) \leq1-\delta
.
\]
Then,
\[
\prob\bigl(\mu(S_n) \leq1\bigr) \rightarrow1
.
\]
\end{lemma}

\begin{pf}
By triangular inequality and thanks to the constraint on $\mu(S)$,
\[
\prob\bigl(\mu(S_n) \leq1\bigr) \geq\prob \bigl(\mu(S_n - S) \leq1 - \mu
(S) \bigr) \geq
\prob\bigl(\mu(S_n - S) \leq\delta\bigr)
,
\]
%
Convergence in probability of $S_n$
to $S$ concludes the proof:
\[
\prob\bigl(\mu(S_n - S) \leq\delta\bigr) \rightarrow1
 \qquad\mbox{therefore }
\prob\bigl(\mu(S_n) \leq1\bigr) \rightarrow1
.
\]
\upqed
\end{pf}

Let us consider the full vector $\tildebbeta^n$ with coefficients
$\tildebbeta_\supp^n$ defined as in Lemma~\ref
{lemsupportconsistency} and
other coefficients null, $\tildebbeta_{\supp^c}^n=\mathbf{0}$.
We now proceed to the last step of the proof of
Theorem~\ref{thmsupportconsistency}, by proving that $\tildebbeta^n$ satisfies
the coop-Lasso optimality conditions with probability tending to 1 under
the additional conditions (A4)--(A5).
The final conclusion then results from the uniqueness of the coop-Lasso
estimator.

First, consider optimality conditions with respect to $\bbeta_{\supp
} $.
As a result of Lemma~\ref{lemsupportconsistency}, the probability that
$\tildebbeta_j^n \neq0$ for every $j \in\supp$ tends to 1.
Thereby, $\tilde{\bbeta}{}_{\supp}^n$ satisfies \eqref{eqoptimality}
on the restriction of $\mathbf{X}$ to covariates in $\supp$ with probability
tending to 1.
As $\tildebbeta_{\supp^c}^n=\mathbf{0}$,
then $\mathbf{X}\tildebbeta^n = \mathbf{X}_{\centerdot\supp
}\tildebbeta_{\supp}^n$
and for every $j \in\supp$,
$\|\bvarphi_j(\tildebbeta_{\supp_k})^n\| = \|\bvarphi_j(\tildebbeta
_{\group}^n)\|$,
therefore, $\tilde{\bbeta}{}_{\supp}^n$ satisfies \eqref{eqoptimality}
in the original problem with probability tending to 1.

Second, $\tildebbeta_{\supp^c}^n$ should also verify the optimality conditions
\eqref{eqoptimalityall} with probability tending to 1.
With assumption (A3), we only have to consider two cases that read:
\begin{itemize}
\item if group $k$ is excluded from the support, one must have
%
%
\begin{eqnarray}\label{eqoptcond2a}
&&\prob \bigl(\max \bigl( \bigl\|
\bigl((\mathbf{X}_{\centerdot\supp_k^c})^\intercal(\mathbf
{X}\tildebbeta^n-\mathbf{y})\bigr)^+
\bigr\|_n, \bigl\|
\bigl((\mathbf{X}_{\centerdot\supp_k^c})^\intercal(\mathbf
{X}\tildebbeta^n-\mathbf{y})\bigr)^-
\bigr\|_n  \bigr) \leq\lambda_n w_k  \bigr) \nonumber
\\[-8pt]
\\[-8pt]&& \qquad \rightarrow1
;
\nonumber\hspace*{-35pt}
\end{eqnarray}
\item if group $k$ intersects the support, with either positive
($\nu_k = 1$) or negative ($\nu_k=-1$) coefficients, one must have
%
\begin{equation}
\prob \bigl( \{\nu_k (\mathbf{X}_{\centerdot\supp_k^c})^\intercal
(\mathbf{X}\tildebbeta^n-\mathbf{y})
\succeq\mathbf{0}\} \cap\{ \|(\mathbf{X}_{\centerdot\supp
_k^c})^\intercal(\mathbf{X}\tildebbeta^n-\mathbf{y})\|_n
\leq\lambda_n w_k \}  \bigr) \rightarrow1
.
\label{eqoptcond2b}\hspace*{-30pt}
\end{equation}
\end{itemize}
To prove \eqref{eqoptcond2a} and \eqref{eqoptcond2b}, we study the
asymptotics of
$(\mathbf{X}_{\centerdot\supp_k^c})^\intercal(\mathbf
{X}\tildebbeta^n-\mathbf{y})/n$
for any group such that $\supp_k^c$ is not empty. As a consequence of
the existence of the fourth order moments of the centered random
variables $X$ and\vadjust{\goodbreak}
$Y$, the multivariate central limit theorem applies, yielding
%
\begin{eqnarray*} \label{eqTCL}
\frac{\mathbf{X}^\intercal\mathbf{X}}{n} &=& \frac{1}{n}
\sum_{i=1}^n \mathbf{x}_{i}^\intercal\mathbf{x}_{i} =
\bPsi+ O_P(n^{-1/2}),\nonumber
\\[-8pt]
\\[-8pt]
\frac{\mathbf{X}^\intercal\boldsymbol{\varepsilon} }{n} &=& \frac{1}{n}
\sum_{i=1}^n \mathbf{x}_{i} \boldsymbol{\varepsilon}_i =
O_P(n^{-1/2})
\nonumber
\end{eqnarray*}
Then, we derive from \eqref{eqTCL} and the definition of $\tildebbeta
^n$ that
%
\begin{eqnarray}\label{eqkeyline1}
\frac{1}{n} (\mathbf{X}_{\centerdot\supp_k^c})^\intercal
(\mathbf{X}\tildebbeta^n-\mathbf{y}) & =&
\frac{1}{n} (\mathbf{X}_{\centerdot\supp_k^c})^\intercal\mathbf{X}
(\tildebbeta^n - \bbeta^\star) -
\frac{1}{n}(\mathbf{X}_{\centerdot\supp_k^c})^\intercal
\boldsymbol\varepsilon\nonumber\\
 & =& \frac{1}{n}(\mathbf{X}_{\centerdot\supp_k^c})^\intercal
\mathbf{X}_{\centerdot\supp}
(\tildebbeta_{\supp}^n - \bbeta_{\supp}^\star) + O_P(n^{-1/2}) \\
& = &\bPsi_{\supp_k^c \supp}(\tildebbeta_{\supp}^n -
\bbeta_{\supp}^\star) + O_P(n^{-1/2})
,\nonumber
\end{eqnarray}
while the combination of \eqref{eqTCL} and optimality conditions
\eqref{eqoptimality} on $\tildebbeta_{\supp}^n$ leads to
%
\begin{equation}
\bPsi_{\supp\supp}(\tildebbeta_{\supp}^n-\bbeta_{\supp}^\star) =
- \lambda_n \mathbf{D}(\tildebbeta_\supp^n)\tildebbeta_\supp^n +
O_P(n^{-1/2})
,\label{eqkeyline2}
\end{equation}
where $\mathbf{D}(\cdot)$ is the weighting matrix \eqref{eqdefD}.
Put \eqref{eqkeyline1} and \eqref{eqkeyline2} together to finally obtain
%
\begin{equation}
\frac{1}{n}(\mathbf{X}_{\centerdot\supp_k^c})^\intercal(\mathbf
{X}\tildebbeta^n-\mathbf{y})
= - \lambda_n \bPsi_{\supp_k^c \supp}
\bPsi_{\supp\supp}^{-1}\mathbf{D}(\tildebbeta_\supp^n)\tildebbeta
_\supp^n +
O_P(n^{-1/2})\hspace*{-30pt}
. \label{eqOp}
\end{equation}

Now, define for any $k$ such that $\supp_k^c$ is not empty:
\[
R_{k,n} = \frac{1}{w_k \lambda_n}\frac{1}{n}
(\mathbf{X}_{\centerdot\supp_k^c})^\intercal(\mathbf{X}\tildebbeta
^n-\mathbf{y})
\quad\mbox{and} \quad R_k = - \frac{1}{w_k} \bPsi_{\supp_k^c \supp}
\bPsi_{\supp\supp}^{-1}\mathbf{D}(\bbeta^\star_\supp)\bbeta
^\star_\supp
.
\]
Limits \eqref{eqoptcond2a} and \eqref{eqoptcond2b} are expressed:
\begin{itemize}
\item if group $k$ is excluded from the support, one must have
\[
\prob \bigl(\max (\|R_{k,n}^+\|,\|R_{k,n}^-\| )\leq
1 \bigr) \rightarrow1
;
\]
\item if group $k$ intersects the support, with either positive
($\nu_k = 1$) or negative ($\nu_k=-1$) coefficients, one must have
\[
\prob \bigl( \{ \nu_kR_{k,n} \succeq\mathbf{0}\} \cap\{\|(\nu
_kR_{k,n})^+\| \leq1\}  \bigr)
\rightarrow1
.
\]
\end{itemize}

Remark that, as a continuous function of $\tildebbeta_\supp^n$,
$\mathbf{D}(\tildebbeta_\supp^n)\tildebbeta_\supp^n$ converges in
probability to
$\mathbf{D}(\bbeta^\star_\supp)\bbeta^\star_\supp$. Therefore,
with a decrease rate
for $\lambda_n$ chosen such that $n^{1/2}\lambda_n
\rightarrow\infty$, equation \eqref{eqOp} implies
%
\begin{equation}
R_{k,n} \inprob R_k . \label{eqconvproba}
\end{equation}

It now suffices to successively apply Lemma \ref
{lemUseIrrepresentableCond} to
the appropriate vectors and norms to show that $\tildebbeta_{\supp
^c}^n$ satisfies
\eqref{eqoptcond2a} and \eqref{eqoptcond2b}:
\begin{itemize}
\item if group $k$ is excluded from the support, (A4) assumes that
there exists
$\eta>0$, such that
\[
\max(\|R_k^+\|,\|R_k^-\|) \leq1 -\eta\vadjust{\goodbreak}
,
\]
and Lemma \ref{lemUseIrrepresentableCond} applied to $\mu(u) =
\max(\|u^+\|,\|u^-\|)$ provides
\[
\prob\{\max(\|R_{k,n}^+\|,\|R_{k,n}^-\|) \leq1\}
\rightarrow1
.
\]
\item if group $k$ intersects the support, with either positive
($\nu_k = 1$) or negative ($\nu_k=-1$) coefficients,
\begin{eqnarray*}
&& \prob \bigl( \{\|(\nu_kR_{k,n})^+\| \leq1\} \cap\{ \nu_kR_{k,n}
\succeq\mathbf{0}\}  \bigr)
\\
& & \qquad   = 1 - \prob
 \bigl(\{\|(\nu_kR_{k,n})^+\| > 1 \} \cup\{ \nu_kR_{k,n} \prec
\mathbf{0} \} \bigr) \\
&& \qquad    \geq1 - \prob
\bigl (\|(\nu_kR_{k,n})^+\| > 1  \bigr) - \prob ( \nu_kR_{k,n}
\prec\mathbf{0} ) \\
&& \qquad    \geq1 - \prob \bigl( \max(\| R_{k,n}^+\|,\|
R_{k,n}^-\|) > 1
 \bigr) - \prob ( \nu_kR_{k,n} \prec\mathbf{0} )
.
\end{eqnarray*}
As previously, the first probability in the sum tends to 0 because of
(A4) and Lemma
\ref{lemUseIrrepresentableCond}. The second
probability tends to 0 from (A5) and of the convergence in probability
of $R_{k,n}$ to $R_k$. Therefore, the overall probability tends to 1.
\end{itemize}

Denote by $A_{k,n}$ these events on which coefficients in $\supp_k^c$
are set to 0. We just showed that individually for each group $k$
with true null coefficients, $P(A_{k,n}) \rightarrow1$. This implies that
\[
\prob \biggl( \bigcup_{k\dvtx  \supp_k^c \neq\varnothing} A^c_{k,n}  \biggr)
\leq\sum_{k\dvtx  \supp_k^c \neq\varnothing} \prob ( A_{k,n}^c
 )
\rightarrow0,
\]
which in turn concludes the proof:
\[
\prob \biggl(\bigcap_{k\dvtx  \supp_k^c \neq\varnothing} A_{k,n}  \biggr)
\rightarrow1.
\]
\end{appendix}

\section*{Acknowledgments}
We would like to thank Marine Jeanmougin for her helpful comments on
the breast cancer data set and for sharing her differential analysis
on the subset of basal tumors. We also thank Catherine Matias for her
careful reading of the manuscript and Christophe Ambroise for fruitful
discussions.


%

\printaddresses

\end{document}